\pdfoutput=1
\documentclass[tikz,11pt,english,a4paper]{article}
\usepackage{float}
\usepackage{setspace}
\usepackage{jcappub}
\usepackage[utf8]{inputenc}
\DeclareUnicodeCharacter{2212}{\ensuremath{-}}
\usepackage{subcaption}
\usepackage{mathtools}

\usepackage{bm}
\usepackage{algorithm}
\usepackage{nicefrac}
\usepackage{bbold}
\usepackage{dsfont}
\usepackage[linewidth=1pt]{mdframed}
\usepackage{bold-extra}
\usepackage{tabto}
\usepackage{amssymb}
\usepackage{booktabs}
\usepackage{multirow}
\usepackage{placeins}

\usepackage[noend]{algpseudocode}
\makeatletter
\def\BState{\State\hskip-\ALG@thistlm}
\makeatother

\usepackage{fontawesome5}
\usepackage{hyperref}
\makeatletter
\newcommand{\github}[1]{%
   \href{#1}{\faGithub}%
}
\makeatother

\newcommand{\connect}{\textsc{connect}}
\newcommand{\client}{CLiENT}
\newcommand{\class}{\textsc{class}}

\newcommand{\montepython}{\textsc{MontePython}}

\newcommand{\cobaya}{Cobaya}
\newcommand{\PDP}{PDP+}
\newcommand{\PDPS}{PDP+SH0ES}
\newcommand{\HWC}{\ensuremath{H_0{\rm WC}}}
\newcommand{\HWCS}{\ensuremath{H_0{\rm WC}}+SH0ES}
\usepackage{siunitx}
\DeclareSIUnit \parsec {pc}

\usepackage[T1]{fontenc}

\usepackage{lmodern} 
\rmfamily 
\DeclareFontShape{T1}{lmr}{b}{sc}{<->ssub*cmr/bx/sc}{}
\DeclareFontShape{T1}{lmr}{bx}{sc}{<->ssub*cmr/bx/sc}{}

\newcommand{\CC}{C\nolinebreak\hspace{-.05em}\raisebox{.4ex}{\tiny\bf +}\nolinebreak\hspace{-.10em}\raisebox{.4ex}{\tiny\bf +}}
\newcommand{\CLASSpp}{\textsc{class}\nolinebreak\hspace{-.05em}\raisebox{.4ex}{\tiny\bf +}\nolinebreak\hspace{-.10em}\raisebox{.4ex}{\tiny\bf +}}

\usepackage{soul}

\usepackage[edges]{forest}
\definecolor{folderbg}{RGB}{124,166,198}
\definecolor{folderborder}{RGB}{110,144,169}
\newlength\Size
\tikzset{%
  folder/.pic={%
    \filldraw [draw=folderborder, top color=folderbg!50, bottom color=folderbg] (-1.05*\Size,0.2\Size+5pt) rectangle ++(.75*\Size,-0.2\Size-5pt);
    \filldraw [draw=folderborder, top color=folderbg!50, bottom color=folderbg] (-1.15*\Size,-\Size) rectangle (1.15*\Size,\Size);
  },
  file/.pic={%
    \filldraw [draw=folderborder, top color=folderbg!5, bottom color=folderbg!10] (-\Size,.4*\Size+5pt) coordinate (a) |- (\Size,-1.2*\Size) coordinate (b) -- ++(0,1.6*\Size) coordinate (c) -- ++(-5pt,5pt) coordinate (d) -- cycle (d) |- (c) ;
  },
}
\forestset{%
  declare autowrapped toks={pic me}{},
  declare boolean register={pic root},
  pic root=0,
  pic dir tree/.style={%
    for tree={%
      folder,
      font=\ttfamily,
      grow'=0,
    },
    before typesetting nodes={%
      for tree={%
        edge label+/.option={pic me},
      },
      if pic root={
        tikz+={
          \pic at ([xshift=\Size].west) {folder};
        },
        align={l}
      }{},
    },
  },
  pic me set/.code n args=2{%
    \forestset{%
      #1/.style={%
        inner xsep=2\Size,
        pic me={pic {#2}},
      }
    }
  },
  pic me set={directory}{folder},
  pic me set={file}{file},
}

\usepackage{newfloat}
\usepackage{caption}
\DeclareFloatingEnvironment[fileext=frm,placement={!t},name=Algorithm]{pseudoenv}
\DeclareCaptionLabelSeparator{normal-colon}{\normalfont : } 
\makeatletter
\g@addto@macro\bfseries{\boldmath}
\makeatother

\makeatletter
\def\old@comma{,}
\catcode`\,=13
\def,{%
  \ifmmode%
    \old@comma\discretionary{}{}{}%
  \else%
    \old@comma%
  \fi%
}
\makeatother

\newcommand{\LCDM}{$\Lambda$CDM}

\newcommand{\Geff}{G_\text{eff}}

\newcounter{meq}

\begin{document}

\title{Self-Interacting Neutrinos in NEDE and DRMD Cosmologies: Posteriors, profiles and evidence using \client{}}

\author[a]{Magnus Uggerhøj,}
\author[a]{Tobias Johannessen,}
\author[a]{Jeppe Pagaard Thybo,}
\author[a]{Steen Hannestad,}
\author[a]{Thomas Tram,}
\affiliation[a]{Department of Physics and Astronomy, Aarhus University,
 DK-8000 Aarhus C, Denmark}

\emailAdd{steen@phys.au.dk}
\emailAdd{thomas.tram@phys.au.dk}

\abstract{A large number of extensions of the $\Lambda$CDM model have been proposed as solutions to the Hubble tension, but they are almost always tested separately. Combinations of such models are rarely studied, and one reason is that their parameter spaces are very costly to explore with standard methods. Here we study a combination of two models which have each been proposed as a solution: self-interacting neutrinos (SI$\nu$) and the dark radiation matter decoupling (DRMD) of hot new early dark energy. The SI$\nu$+DRMD model has twelve cosmological parameters, a bimodal posterior in the neutrino self-interaction and several flat directions in parameter space, and it can only be investigated in detail because we use the likelihood emulator \client{}. With emulators trained on about $3\times10^5$ likelihood evaluations per dataset we obtain posteriors, best fits and profile likelihoods for the model and for the models nested within it, and all quoted $\chi^2$ values are calculated with the true likelihood. We use two data combinations, both with and without the SH0ES calibration: a conservative one consisting of Planck, DESI BAO and Pantheon+ supernovae, and a more aggressive one in which the small-scale CMB spectra come from ACT and SPT. With the most aggressive combination, i.e.\ with the small-scale CMB data and SH0ES, the SI$\nu$+DRMD model provides a much better fit than either SI$\nu$ or DRMD on its own, and its posterior gives $H_0 = 70.94\pm 0.60$~km\,s$^{-1}$\,Mpc$^{-1}$. It lowers the $\chi^2$ of $\Lambda$CDM by 27.6, compared with 9.9 for DRMD and 4.0 for SI$\nu$, i.e.\ by more than the sum of the two ($\Delta$AIC $= -15.7$ relative to DRMD, in terms of the Akaike information criterion). This shows that combined models can fit the data considerably better than the individual models, and it demonstrates the value of being able to analyse such complex models in detail. In the other three cases the improvement over the individual models is smaller, and once the penalty for the additional parameters is taken into account the AIC does not significantly prefer the combination.}

\maketitle

\section{Introduction}
\label{sec:introduction}

As the precision of cosmological measurements has steadily increased over the last decade, a number of tensions and possible cracks have started to appear in the framework of the standard $\Lambda$CDM model \cite{Verde:2019ivm,DiValentino:2021izs,Abdalla:2022yfr,CosmoVerseNetwork:2025alb}. The most often discussed of these is the Hubble tension, a rather strong discrepancy between the value of $H_0$ directly measured from the distance ladder and type Ia supernovae \cite{Riess:2021jrx} and the value inferred from CMB and large-scale structure data within the context of the $\Lambda$CDM model \cite{Planck:2018vyg}. However, other tensions have recently become apparent, such as the preference for a negative neutrino mass in the combined CMB and BAO data from Planck and DESI \cite{Craig:2024tky,Green:2024xbb,DESI:2025zgx,Elbers:2025vlz}.

While missing systematics in the data remain a viable explanation for some or all of these discrepancies, the available evidence is pointing to a modification of the simplest $\Lambda$CDM model as the most likely outcome.
Many possible extensions of the cosmological standard model have been investigated \cite{DiValentino:2021izs,Abdalla:2022yfr,CosmoVerseNetwork:2025alb}, and by now it seems clear that any viable model attempting to explain e.g.\ the Hubble tension should preferably modify physics around the epoch of matter--radiation equality and photon decoupling \cite{Bernal:2016gxb,Aylor:2018drw,Knox:2019rjx}.
Models which can achieve this are for example self-interacting neutrinos with non-standard interactions mediated by either massive \cite{Cyr-Racine:2013jua,Lancaster:2017ksf,Oldengott:2017fhy,Kreisch:2019yzn,Blinov:2019gcj,RoyChoudhury:2020dmd,Brinckmann:2020bcn,Barenboim:2026jef} or massless \cite{Hannestad:2004qu,Bell:2005dr,Basboll:2008fx,Archidiacono:2013dua,Forastieri:2015paa} scalars or pseudo-scalars, and models which postulate a period around matter--radiation equality in which a significant fraction of the energy density is in a potential-dominated scalar field, i.e.\ early dark energy \cite{Poulin:2023lkg,Kamionkowski:2022pkx}. The latter class includes the hot new early dark energy model \cite{Niedermann:2021ijp,Niedermann:2021vgd,Garny:2024ums} and its dark radiation matter decoupling \cite{Garny:2025kqj}, which provides the dark sector we consider here.
Both these types of models have been thoroughly investigated in many papers over the past decade, as have other early-time solutions to the Hubble tension. In the case of self-interacting neutrinos some CMB and large-scale structure data shows a preference for a strongly interacting mode \cite{Kreisch:2019yzn,Kreisch:2022zxp,Camarena:2023cku,He:2023oke}, but this preference has not been confirmed by other large-scale structure data \cite{Camarena:2024daj,Poudou:2025qcx,He:2025jwp} or by the small-scale CMB measurements of ACT and SPT \cite{AtacamaCosmologyTelescope:2025nti,Noriega:2026uhw}. Several systematic comparisons of the proposed solutions using common datasets have also been carried out \cite{Schoneberg:2021qvd,Escudero:2022rbq,Khalife:2023qbu,Schoneberg:2026eys,Schoneberg:2026vaf}. In the most recent of these, the $H_0$ World Cup, even the best performing models leave a residual discrepancy of about 2.5--3.6$\sigma$ with the SH0ES measurement \cite{Schoneberg:2026eys}. Self-interacting dark radiation and DRMD were among the models tested, whereas self-interacting neutrinos were not included because they are already strongly constrained \cite{Schoneberg:2026vaf}.

However, studies in which two or more of these models have been combined are much less common. It has been argued that a single early-time mechanism may not be enough to resolve the tension \cite{Vagnozzi:2023nrq}, and models with more than one new ingredient have been proposed in order to address several tensions at once \cite{Allali:2021azp,Clark:2021hlo}. Early dark energy has also been combined with massive neutrinos \cite{Reeves:2022aoi}, and the latest systematic comparison tests whether late-time extensions can improve the early-time solutions \cite{Schoneberg:2026vaf}. One reason why such combinations are rarely studied is of course the inherent complexity in the underlying model framework. However, another important reason is that such models are computationally extremely costly to investigate using standard tools of cosmological parameter estimation. Combined models typically exhibit many parameter degeneracies and highly non-Gaussian directions in parameter space and therefore require a very large number of likelihood evaluations in order to calculate posterior distributions for the involved cosmological parameters. Emulators of the observables \cite{SpurioMancini:2021ppk,Nygaard:2022wri,Bonici:2023xjk,Gunther:2025xrq} or of the posterior and the likelihood \cite{Gammal:2022eob,Cohen:2026iij} can reduce this cost very substantially.

In this work we present a worked example of such a combination. We combine two models which have each been proposed as a solution to the Hubble tension, self-interacting neutrinos and the DRMD model, and ask whether the combined parameter space contains better solutions than either model on its own once the additional parameters are taken into account. In order to do this we use the likelihood emulator \client{} \cite{Janken:2025wlq}\footnote{\url{https://github.com/AarhusCosmology/client-f}}, which learns the likelihood of the data directly as a function of all model and nuisance parameters using a neural network trained by active learning. We analyse the SI$\nu$+DRMD model and the models nested within it using two different data combinations, both with and without the SH0ES calibration: the full Planck temperature, polarisation and lensing data together with DESI DR2 BAO and Pantheon+ supernovae (Planck+DESI+Pantheon+, or \PDP{}), and the data combination of the $H_0$ World Cup \cite{Schoneberg:2026eys,Schoneberg:2026vaf} (\HWC{}), in which the small angular scales of the CMB are measured by ACT DR6 and SPT-3G. Apart from posteriors the emulators also make it practical to calculate the Bayesian evidence and to compute profile likelihoods with the true likelihood. Profile likelihoods are particularly useful here because they separate what the data prefers from the effects of prior volume \cite{Herold:2021ksg,Gomez-Valent:2022hkb,Holm:2022kkd,Herold:2024enb}, which are large in a parameter space with flat directions and a coupling that spans many orders of magnitude. The evidence likewise depends strongly on the choice of prior \cite{Trotta:2008qt}, and we therefore compare models using their best fits and the Akaike information criterion. We find that with \HWC{} and the SH0ES calibration the SI$\nu$+DRMD model fits the data substantially better than either SI$\nu$ or DRMD on its own, also after the penalty for the additional parameters has been taken into account. At the same time the small-scale CMB data largely removes the strongly interacting mode of the neutrino self-interaction.

The paper is organised as follows. In Section~\ref{sec:theory} we describe the model framework and in Sec.~\ref{sec:numerics} the numerical setup, including the data. Section~\ref{sec:results} contains our results and Sec.~\ref{sec:conclusions} our discussion and conclusions. Finally, the emulators and their accuracy are described in detail in Appendix~\ref{app:client}, and the computation of the best fits and the profile likelihoods in Appendix~\ref{app:bestfits}.

\section{Model framework}
\label{sec:theory}

The cosmological model we investigate here contains two distinct pieces of physics beyond the standard $\Lambda$CDM cosmology: self-interactions in the neutrino sector, and a new interacting dark sector.

\subsection{Self-interacting neutrinos}
\label{sec:theory:sinu}

New interactions in the neutrino sector have a phenomenological impact which is highly dependent on specifics of the underlying physical model. The simplest (and by far the most often studied) model is one in which all neutrinos are charged under a new interaction mediated by a massive scalar or pseudo-scalar. Provided that the mass of the mediator is significantly higher than all other energy scales relevant to the problem it can be assumed that any pre-existing population of the mediator particles has annihilated away and that no physical mediator particles can subsequently be produced.
Depending on whether this annihilation process takes place before or after the standard-model neutrino decoupling at $T \sim 1$ MeV this may lead to a change in the total neutrino energy density. However, this can be absorbed into the parameter $N_{\rm eff}$ which characterises the total relativistic energy density in the neutrino sector.
This reduces the problem to one in which neutrinos are subject only to elastic scatterings mediated by the new interaction. In such a setup the background evolution is completely unchanged, but perturbations are changed because the new interaction introduces a term which dampens neutrino free-streaming and leads to suppression of anisotropic stress in the neutrino sector. Since free-streaming neutrinos shift the phase of the acoustic oscillations and suppress their amplitude \cite{Bashinsky:2003tk,Baumann:2015rya,Follin:2015hya}, a delay of the onset of free-streaming changes both.

In order to follow the evolution of perturbations in the neutrino sector it is necessary to solve the collisional Boltzmann equation. Here we follow the same notation as in \class{} (which in turn uses the notation presented in \cite{Ma:1995ey}).

The Boltzmann equation for the evolution of a given species can generically be written as
\begin{equation}
L[f] = \frac{Df}{D\tau} = C[f],
\end{equation}
where $L[f]$ is the Liouville operator, $C[f]$ is the collision operator,  and we use conformal time, defined as $d \tau = dt/a(t)$, as our time variable.

Following \cite{Ma:1995ey}  we write the distribution function as
\begin{equation}
f(x^i,q,n_j,\tau) = f_0(q) [1+\Psi(x^i,q,n_j,\tau)],
\end{equation}
where $f_0(q)$ is the unperturbed distribution function and $\Psi(x^i,q,n_j,\tau)$ is the perturbation.
As the momentum variable we use the
comoving momentum $q_j \equiv a p_j$. We further parametrise $q_j$ as
$q_j = q n_j$, where $q$ is the magnitude of the comoving momentum and $n_j$ is a unit 3-vector specifying direction.

In synchronous gauge the first-order perturbed Boltzmann equation can be written as an evolution equation for $\Psi$ in $k$-space \cite{Ma:1995ey}
\begin{equation}
\frac{1}{f_0} L[f] = \frac{\partial \Psi}{\partial \tau} + i \frac{q k}{\epsilon}
\mu \Psi + \frac{d \ln f_0}{d \ln q} \left[\dot{\eta}-\frac{\dot{h}+6\dot{\eta}}
{2} \mu^2 \right] = \frac{1}{f_0} C[f],
\label{eq:boltzX}
\end{equation}
where $\mu \equiv n^j \hat{k}_j$ and $\epsilon = (q^2+a^2 m^2)^{1/2}$.
$h$ and $\eta$ are the metric perturbations, defined from the perturbed space-time
metric in synchronous gauge~\cite{Ma:1995ey}
\begin{align}
ds^2 &= a^2(\tau) [-d\tau^2 + (\delta_{ij} + h_{ij})dx^i dx^j], \\
h_{ij} &= \int d^3 k e^{i \vec{k}\cdot\vec{x}}\left(\hat{k}_i \hat{k}_j h(\vec{k},\tau)
+(\hat{k}_i \hat{k}_j - \frac{1}{3} \delta_{ij}) 6 \eta (\vec{k},\tau) \right).
\end{align}

Performing a Legendre expansion of the perturbation, $\Psi = \sum_\ell (-i)^\ell (2\ell+1)\, \Psi_\ell\, P_\ell(\mu)$, then leads to the usual hierarchy of equations for the multipoles $\Psi_\ell$ \cite{Ma:1995ey}, and we can perform the corresponding expansion of the collision operator in order to derive the full set of hierarchy equations to be solved numerically.
For the interaction we investigate here, the full collision operator has been calculated in \cite{Oldengott:2014qra,Oldengott:2017fhy}. However, here we will use the numerically much simpler relaxation time approximation (RTA) \cite{Hannestad:2000gt,Hannestad:2004qu} in which the collision term reduces to a simple and momentum-independent damping rate of all multipoles beyond $\ell=1$ (energy and momentum conservation in $\nu \nu \rightarrow \nu \nu$ scattering protect the monopole and dipole, so the RTA collision term vanishes for $\ell = 0,1$ and acts only on $\ell \geq 2$).

In synchronous gauge the collisional Boltzmann hierarchy for massive interacting neutrinos is then given by \cite{RoyChoudhury:2020dmd,RoyChoudhury:2022rva}
\begin{align}
    \dot{\Psi}_0
    &=
    -\frac{qk}{\epsilon}\Psi_1
    + \frac{1}{6}\dot{h}
    \frac{\mathrm{d}\ln f_0}{\mathrm{d}\ln q},
    \\
    \dot{\Psi}_1
    &=
    \frac{qk}{3\epsilon}
    \left(\Psi_0 - 2\Psi_2\right),
    \\
    \dot{\Psi}_2
    &=
    \frac{qk}{5\epsilon}
    \left(2\Psi_1 - 3\Psi_3\right)
    -
    \left(
        \frac{1}{15}\dot{h}
        + \frac{2}{5}\dot{\eta}
    \right)
    \frac{\mathrm{d}\ln f_0}{\mathrm{d}\ln q}
    - \alpha_2 \dot{\tau}_\nu \Psi_2,
    \\
    \dot{\Psi}_\ell
    &=
    \frac{qk}{(2\ell+1)\epsilon}
    \left[
        \ell\Psi_{\ell-1}
        -(\ell+1)\Psi_{\ell+1}
    \right]
    - \alpha_\ell \dot{\tau}_\nu \Psi_\ell,
    \qquad \ell \geq 3,
\end{align}
where
\begin{equation}
    \dot{\tau}_\nu \equiv a\, G_\text{eff}^2\, T_\nu^5 \propto a^{-4}
\end{equation}
is the conformal time opacity of the self-interaction, $T_\nu$ is the neutrino temperature, and the dimensionless coefficients $\alpha_\ell$ encode the angular structure of the matrix element. Note that refs.~\cite{Cyr-Racine:2013jua,Lancaster:2017ksf} define the opacity with the opposite sign and therefore write the damping terms as $+\alpha_\ell \dot{\tau}_\nu \Psi_\ell$. In the absence of the damping terms these are simply the equations for collisionless massive neutrinos \cite{Ma:1995ey}. The fluid variables $\delta_\nu$, $\theta_\nu$ and $\sigma_\nu$ are recovered as momentum integrals over $\Psi_0$, $\Psi_1$ and $\Psi_2$ weighted respectively by $\epsilon$, $q$ and $q^2/\epsilon$ \cite{Ma:1995ey}.

We use the coefficients calculated in ref.~\cite{Oldengott:2017fhy} for a flavour-independent scalar interaction, $\alpha_2 = 0.40$, $\alpha_3 = 0.43$, $\alpha_4 = 0.46$, $\alpha_5 = 0.47$ and $\alpha_{\ell \geq 6} = 0.48$, as was also done in refs.~\cite{RoyChoudhury:2020dmd,RoyChoudhury:2022rva}. These coefficients have recently been calculated fully analytically within the same approximation \cite{Mogensen:2026vqj}, and the exact values are 4--6\% larger. However, this difference is almost completely degenerate with $\Geff$ \cite{Mogensen:2026vqj} because the damping is proportional to $\alpha_\ell \Geff^2$, and it corresponds to a shift of $\log_{10}(\Geff\,\text{MeV}^2)$ by about $-0.013$, far below the uncertainties found here. Both calculations use Maxwell--Boltzmann statistics for the neutrinos. The coefficients have also been calculated with Fermi--Dirac statistics and Pauli blocking, but with a different normalisation of $\alpha_\ell$ \cite{Perez-Castro:2026muj}, and we do not include the effect of quantum statistics here. If only some of the neutrino flavours interact, a consistent treatment of neutrino mixing shows that bounds on $\Geff$ derived for a flavour-independent interaction are weakened by a factor of $\sqrt{2}$ when two flavours interact and $\sqrt{6}$ when only one does \cite{Barenboim:2026jef}.
The framework extends straightforwardly to a species-specific scenario in which the damping term is applied only to a subset of the mass eigenstates \cite{Das:2020xke,RoyChoudhury:2022rva}. However, here we assume that all three neutrinos interact with a common $\Geff$.

The phenomenologically interesting range of $\Geff$ is dictated by the epoch at which one wants the neutrinos to free-stream. The standard weak interaction, $\Geff = G_\text{F}$, leads to free-streaming for $T \lesssim \text{MeV}$, while couplings of order $\Geff \sim 10^9 G_\text{F}$ delay the onset of free-streaming until $z \sim 10^4$, shortly before matter--radiation equality, and produce observable effects in the CMB \cite{Cyr-Racine:2013jua,Lancaster:2017ksf,Kreisch:2019yzn,RoyChoudhury:2022rva}.

Previous analyses using Planck data have found a bimodal posterior in $\log_{10}(\Geff\,\text{MeV}^2)$, with a moderately interacting mode around $\log_{10}(\Geff\,\text{MeV}^2) \sim -4$, which extends all the way to the non-interacting limit, and a strongly interacting mode near $\log_{10}(\Geff\,\text{MeV}^2) \sim -1.7$ \cite{Kreisch:2019yzn,RoyChoudhury:2020dmd,RoyChoudhury:2022rva}. The strongly interacting mode is allowed by CMB data because of a degeneracy between $\Geff$, the angular size of the sound horizon $\theta_s$, the scalar spectral index $n_s$ and the amplitude $A_s$. In this mode the neutrinos begin to free-stream only at $z \approx 8000$, roughly when modes corresponding to multipoles $\ell \approx 400$ enter the horizon \cite{Kreisch:2019yzn}. In this work we adopt a flat prior on $\log_{10}(\Geff\,\text{MeV}^2)$, motivated by the very wide range of physically interesting values, with the specific prior bounds discussed in Sec.~\ref{subsec:param_and_prios}.

The competition between $\dot{\tau}_\nu$ and the comoving wavenumber $k$ determines the behaviour of the neutrinos for each Fourier mode \cite{Oldengott:2017fhy}. In the strong-coupling regime, $\dot{\tau}_\nu \gg k$, the multipoles $\ell \geq 2$ are damped away and the neutrinos behave as a perfect fluid without anisotropic stress. In the weak-coupling regime, $\dot{\tau}_\nu \ll k$, the damping is negligible and the standard collisionless cascade populates the higher multipoles, generating $\sigma_\nu \neq 0$ and the usual splitting of the gravitational potentials. The transition, $\dot{\tau}_\nu \sim k$, occurs at a definite epoch for each mode, and since $\dot{\tau}_\nu \propto a^{-4}$ while $k$ is constant, the effect of the neutrinos differs between modes which enter the horizon at different times. The modification of the CMB and matter power spectra is therefore intrinsically scale dependent, and this makes it possible to constrain the strength of the self-interaction observationally \cite{Kreisch:2019yzn,RoyChoudhury:2022rva}.

Because $\dot{\tau}_\nu/\mathcal{H} \propto a^{-3}$ during radiation domination the hierarchy becomes numerically stiff at early times, and earlier implementations have therefore used a tight-coupling approximation in this regime \cite{RoyChoudhury:2020dmd,RoyChoudhury:2022rva}. \CLASSpp{} instead integrates the full hierarchy using an implicit (stiff) solver and caps the opacity at $\dot{\tau}_\nu = 10^9 \mathcal{H}$. At such high rates the multipoles $\ell \geq 2$ are damped to completely negligible amplitudes, and the cap therefore does not affect the solution.

A complementary simplification is made deep inside the Hubble radius. Once $k\tau$ exceeds a threshold (31 by default and 15 with the precision settings for ACT DR6 of Sec.~\ref{sec:numerics:classpp}), \CLASSpp{} replaces the hierarchy of the massive neutrinos by the fluid approximation of ref.~\cite{Lesgourgues:2011rh}, which retains only the density contrast $\delta_\nu$, the velocity divergence $\theta_\nu$ and the shear $\sigma_\nu$. This approximation is accurate in particular because the free-streaming term $qk/\epsilon$ decreases once the neutrinos become non-relativistic, $\epsilon = (q^2 + a^2 m_\nu^2)^{1/2} \rightarrow a m_\nu \gg q$, so that the cascade to higher multipoles stalls. In synchronous gauge, and including the self-interaction, the fluid equations are \cite{Lesgourgues:2011rh}
\begin{align}
    \dot{\delta}_\nu
    &=
    -(1+w_\nu)
    \left(
        \theta_\nu + \frac{\dot{h}}{2}
    \right)
    -3\mathcal{H}
    \left(
        c_{\mathrm{eff}}^2 - w_\nu
    \right)
    \delta_\nu,
    \\
    \dot{\theta}_\nu
    &=
    -\mathcal{H}
    \left(
        1 - 3c_a^2
    \right)
    \theta_\nu
    +
    \frac{c_{\mathrm{eff}}^2}{1+w_\nu}
    k^2\delta_\nu
    -
    k^2\sigma_\nu,
    \\
    \dot{\sigma}_\nu
    &=
    -3
    \left[
        \frac{1}{\tau}
        +
        \mathcal{H}
        \left(
            \frac{2}{3} - c_a^2 - \frac{\mathfrak{p}_\nu}{3p_\nu}
        \right)
    \right]
    \sigma_\nu
    +
    \frac{4}{3}
    \frac{c_{\mathrm{vis}}^2}{1+w_\nu}
    \left(
        2\theta_\nu + \dot{h}
    \right)
    -
    \alpha_2 \dot{\tau}_\nu \sigma_\nu ,
\end{align}
where $w_\nu = p_\nu/\rho_\nu$ is the equation of state, $\mathfrak{p}_\nu$ the pseudo-pressure \cite{Lesgourgues:2011rh}, $c_a^2 = w_\nu (5 - \mathfrak{p}_\nu/p_\nu)/[3(1+w_\nu)]$ the adiabatic sound speed, and the effective sound speed and the viscosity speed \cite{Hu:1998kj} which close the system are set to $c_\text{eff}^2 = c_a^2$ and $c_\text{vis}^2 = 3 w_\nu c_a^2$. As required by energy and momentum conservation the continuity and Euler equations are unchanged by the self-interaction, and the entire effect of the interaction is contained in the single additional damping term $-\alpha_2 \dot{\tau}_\nu \sigma_\nu$ in the shear equation. Physically, this is the natural extension of the perfect-fluid behaviour seen in the strong-coupling limit of the full hierarchy: when the shear is driven to zero, the anisotropic stress vanishes and the neutrinos again behave as a perfect fluid.

The interacting neutrino sector therefore enters the Einstein--Boltzmann system only through the damping of the higher neutrino multipoles. The resulting change in the anisotropic stress affects the rest of the system, i.e.\ the Einstein equations, the photon hierarchy, the baryon fluid equations etc., only indirectly through the gravitational potentials sourced by the modified $\sigma_\nu$. Together with the hot NEDE/DRMD dark sector discussed in Secs.~\ref{sec:theory:HotNEDE} and \ref{sec:theory:DRMD} the model contains two independent sectors in which an interaction rate is compared with the expansion rate or with the wavenumber of a mode, and both modify the gravitational potentials. Note, however, that the two sectors do not interact directly with each other and are coupled only through gravity.

\subsection{Hot New Early Dark Energy}
\label{sec:theory:HotNEDE}

The hot new early dark energy (hot NEDE) model was introduced by Niedermann and Sloth~\cite{Niedermann:2021ijp,Niedermann:2021vgd} as an extension of new early dark energy \cite{Niedermann:2019olb,Niedermann:2020dwg}, and has subsequently been developed further in refs.~\cite{Garny:2024ums,Garny:2025kqj}. The model describes an interacting dark sector consisting of dark radiation and dark matter, with self-interactions in the dark radiation sector as well as interactions between the dark radiation and a fraction of the dark matter. The underlying particle physics realisation is based on a dark non-Abelian gauge theory in which $SU(N)$ is broken to $SU(N-1)$, and we refer to refs.~\cite{Garny:2024ums,Garny:2025kqj} for a detailed description.

A defining feature of the model is a supercooled first-order phase transition which takes place after Big Bang Nucleosynthesis, but before recombination. During this transition the vacuum energy stored in the dark sector is converted into a thermal bath of self-interacting dark radiation. This produces a step-like increase in the effective number of relativistic species, $\Delta N_{\mathrm{eff}}$, while automatically satisfying BBN constraints. The resulting increase in the expansion rate prior to recombination reduces the sound horizon and thereby alleviates the Hubble tension \cite{Garny:2024ums}. In this way hot NEDE provides a post-BBN origin for the stepped self-interacting dark radiation proposed in refs.~\cite{Aloni:2021eaq,Schoneberg:2022grr}.

The same phase transition also generates a small mass splitting between different dark matter states. At early times the dark matter and the dark radiation are tightly coupled, but as the dark sector cools the heavier interacting states become Boltzmann suppressed, and the coupling weakens and eventually switches off. This dynamical decoupling mechanism is known as dark radiation matter decoupling (DRMD), and it is described in more detail in the next subsection. Models in which a fraction of the dark matter interacts with dark radiation, also via an interaction which switches off as the dark radiation cools, have previously been studied in refs.~\cite{Buen-Abad:2015ova,Lesgourgues:2015wza,Buen-Abad:2017gxg,Rubira:2022xhb,Joseph:2022jsf,Buen-Abad:2022kgf,Buen-Abad:2023uva}.

In this work we adopt the large-$N$ limit of the model developed in ref.~\cite{Garny:2025kqj}. For phase transition redshifts in the range $10^6 \lesssim z_* \lesssim 10^9$ all CMB and large-scale structure modes of observational interest remain outside the horizon during the transition. The phase transition itself can therefore be integrated out, and standard adiabatic initial conditions can be imposed. In the large-$N$ limit the small second step in $\Delta N_\mathrm{eff}$, which occurs when the dark Higgs boson becomes non-relativistic, is negligible as well. The resulting cosmological model is then fully specified by four parameters beyond \LCDM:
\begin{equation}
\left\{\Delta N_{\mathrm{eff}}, f_{\mathrm{idm}}, z_{\mathrm{stop}},(\mathcal{G}/\mathcal{H})_{\mathrm{ini}}  \right\},
\end{equation}
where $\Delta N_{\mathrm{eff}}$ is the dark radiation after the transition, $f_{\mathrm{idm}}$ is the fraction of the dark matter which interacts with the dark radiation fluid, and $z_\mathrm{stop}$ and $(\mathcal{G}/\mathcal{H})_\mathrm{ini}$ describe the interaction (see Sec.~\ref{sec:theory:DRMD}). In the limit $f_\text{idm} \rightarrow 0$ the model reduces to self-interacting dark radiation (SIDR) \cite{Jeong:2013eza}.

\subsection{Dark Radiation Matter Decoupling}
\label{sec:theory:DRMD}

Reducing the sound horizon by adding extra radiation does not in itself resolve the Hubble tension. As emphasised in ref.~\cite{Garny:2025kqj}, raising $H_0$ in this way requires an increase of $\omega_m = \Omega_m h^2$ by 10--15\%, which advances matter--radiation equality and enhances the gravitational potentials on scales entering the horizon around equality. With self-interacting dark radiation alone this limits how much $H_0$ can be increased without degrading the fit to the CMB \cite{Garny:2025kqj}. The role of the DRMD mechanism is to provide a correction at the perturbation level which partly compensates this effect, and this is achieved through an interaction which is dynamical, i.e.\ strong at early times and switching off as the dark sector cools.

After the phase transition the charged dark matter states $\chi_j$ interact with the massless dark gauge bosons through a non-Abelian Compton-like process, $\chi_j A^a \rightarrow \chi_j A^a$ \cite{Buen-Abad:2015ova,Rubira:2022xhb}. The resulting momentum transfer rate of a charged state is given by (in cosmic time) \cite{Buen-Abad:2015ova,Rubira:2022xhb,Garny:2025kqj}
\begin{equation}
    \Gamma_{\chi_j A_\mu^a}= \frac{\pi \alpha_d^2}{18 m_\chi} \eta_d T_d^2 [\ln(\alpha_d^{-1})+d_0],
\end{equation}
where $\alpha_d = g^2 /(4 \pi)$ is the dark fine-structure constant, $\eta_d = 2((N-1)^2 -1)$ is the number of massless gauge boson degrees of freedom, and $d_0 \simeq -1.34 - \ln(N-1)$. Since $T_d \propto a^{-1}$ and $H \propto a^{-2}$ the ratio $\Gamma_{\chi_j A_\mu^a}/H$ is approximately constant during radiation domination and decreases only slowly afterwards. Therefore the scattering alone cannot end the coupling.

The decoupling is instead caused by the radiatively induced mass splitting, $\Delta m$, between the charged and the neutral states. As long as $T_d \gg \Delta m$ equipartition holds and the multiplet is populated according to the multiplicity, $f_{\chi_j}/f_\chi = (N-1)/N$. However, once $T_d \ll \Delta m $ the heavier charged states become Boltzmann suppressed, and in equilibrium \cite{Garny:2025kqj}
\begin{equation}
    \left(\frac{f_{\chi_j}}{f_\chi} \right)_\text{eq} \xrightarrow{T_d \ll \Delta m}
    (N-1)\mathrm{e}^{-\Delta m/T_d}.
\end{equation}
Because only the charged states scatter efficiently with the dark radiation the effective drag rate of the interacting dark matter, averaged over its charged and neutral states, is
\begin{equation}
    \Gamma_\text{drag} = \frac{f_{\chi_j}}{f_\chi} \Gamma_{\chi_j A_\mu^a},
\end{equation}
which drops exponentially once the Boltzmann suppression sets in. This defines two characteristic redshifts: $z_\text{stop}$, where $T_d = \Delta m$ and the exponential suppression begins, and $z_\text{dec}$, where $\Gamma_\text{drag}= H$ and the tight coupling breaks down. In the microscopic model the two are related by $(1+z_\text{stop})/(1+z_\text{dec})\simeq 30$ over a wide range of parameters. Together with the hierarchy of mass scales the requirement $z_\text{dec} \sim 10^3$ then leads to the lower end, $z_* \gtrsim 10^6$, of the window of phase transition redshifts, while the upper end of the window is set by BBN \cite{Garny:2025kqj}.

At the perturbation level the dark sector is described by two additional fluids added to the \LCDM{} system: dark radiation (DR) with $w_\text{DR} = c_s^2 = 1/3$ and vanishing anisotropic stress (justified by the strong non-Abelian self-interactions), and the interacting dark matter ($\chi$). In synchronous gauge their equations are \cite{Buen-Abad:2017gxg,Garny:2025kqj}
\begin{gather}
    \delta_\chi'
    + \theta_\chi
    + \frac{h'}{2}
    = 0,
    \qquad
    \theta_\chi'
    + \mathcal{H}\theta_\chi
    = \mathcal{G}\left(\theta_\mathrm{DR} - \theta_\chi\right),
    \\
    \delta_{\mathrm{DR}}'
    + \frac{4}{3}
    \left(
        \theta_{\mathrm{DR}}
        + \frac{h'}{2}
    \right)
    = 0,
    \qquad
    \theta_{\mathrm{DR}}'
    - \frac{k^2}{4}
    \delta_{\mathrm{DR}}
    = -\mathcal{G}R\left(\theta_\mathrm{DR} - \theta_\chi\right),
\end{gather}
where primes denote derivatives with respect to conformal time, $\mathcal{H} = aH$, $\theta_\text{DR}-\theta_\chi$ is the velocity slip between the two fluids, $R = 3\rho_\chi/(4\rho_\text{DR})$ measures their relative inertia, and $\mathcal{G} = a\Gamma_\text{drag}$ is the conformal drag rate, so that $\mathcal{G}/\mathcal{H} = \Gamma_\text{drag}/H$. The structure is directly analogous to the baryon--photon system, with the dark matter playing the role of the baryons and the dark radiation playing the role of the photons. When $\mathcal{G}/\mathcal{H} \gg 1$ the velocity slip is driven to zero and the two fluids share a common velocity (the tight-coupling regime), whereas in the opposite limit the two fluids evolve independently. The tightly coupled fluid has the sound speed $c_s^2 = [3(1+R)]^{-1}$. When the dark radiation is scarce compared with the interacting dark matter, $R \gg 1$, this fraction of the dark matter therefore behaves as a fluid with a small sound speed, which suppresses the growth of structure below its sound horizon until decoupling. For very small $\Delta N_\mathrm{eff}$ and a large $f_\mathrm{idm}$ the predictions then depend on non-linear scales which current modelling cannot describe reliably, and we therefore exclude this region from the analysis (see Sec.~\ref{subsec:param_and_prios}).

The time dependence of $\mathcal{G}$ follows the $T_d^2$ scaling of the rate and the Boltzmann suppression of the charged states. Following ref.~\cite{Garny:2025kqj} it is modelled at all redshifts as
\begin{equation}
    \mathcal{G} = \mathcal{H}
    \left(\frac{\mathcal{G}}{\mathcal{H}}\right)_{\mathrm{ini}}
    \left(
        1 + \frac{\Omega_m}{\Omega_{\mathrm{rad}}}
        \frac{1}{1+z}
    \right)^{-1/2}
    \exp\left(
        -\frac{1+z_{\mathrm{stop}}}{1+z}
    \right),
\end{equation}
where $(\mathcal{G}/\mathcal{H})_\text{ini}$ is the ratio at early times when the coupling is tight, and the first factors keep $\Gamma/H$ constant during radiation domination. The last factor suppresses the interaction exponentially once $1+z$ falls below $1+z_\text{stop}$. The system starts out strongly coupled and decouples, $\mathcal{G} = \mathcal{H}$, at $(1+z_\text{stop})/(1+z_\text{dec}) \approx \ln(\mathcal{G}/\mathcal{H})_\text{ini}$ if this happens during radiation domination. This means that $z_\text{stop}$ and $(\mathcal{G}/\mathcal{H})_\text{ini}$ are largely degenerate, and that the data mainly constrains $z_\text{dec}$ \cite{Garny:2025kqj}. Within our prior (Table~\ref{tab:priors}) $\ln(\mathcal{G}/\mathcal{H})_\text{ini}$ varies from about 5 to 32.

The main effect at the perturbation level is a scale-dependent modification of the Weyl potential. Modes which enter the horizon after dark matter decoupling $(k \lesssim \mathcal{H}(z_\text{dec}))$ evolve as in pure SIDR with no interacting dark matter, whereas modes which enter while the dark matter is still coupled $(k \gtrsim \mathcal{H}(z_\text{dec}))$ experience a drag which suppresses the growth of $ \delta_\chi$ and induces additional decay of the gravitational potential \cite{Garny:2025kqj}. The compensation is most effective when $z_\text{dec} \sim z_\text{eq}$ because the affected modes are then precisely the ones whose evolution is modified by the larger $\omega_m$. Using Planck 2018, DESI DR2, Pantheon+ and the SH0ES calibration, the DRMD analysis of ref.~\cite{Garny:2025kqj} finds a best fit with $z_\text{dec}\approx 2200$ and $f_\text{idm} \approx 3\%$, and an upper bound of $f_\text{idm} < 4.4 \%$ at 95\%~CL.

\section{Numerical setup}
\label{sec:numerics}

\subsection{\CLASSpp{}}
\label{sec:numerics:classpp}

The theoretical model described in the previous section, i.e.\ SI$\nu$ combined with hot NEDE and DRMD, has been implemented in the Boltzmann solver \CLASSpp{}, a \CC{} version of \class{} \cite{Lesgourgues:2011re,Blas:2011rf}, and the two sectors have been thoroughly tested against previous separate implementations of SI$\nu$ and hot NEDE in \class{}. However, since the SI$\nu$+DRMD model has not been investigated before there is no existing code against which the full joint computation can be compared. We have therefore checked each sector separately against an implementation written by a different author in a different code base, so that any feature of the combined result can be attributed to the interplay of the two extensions and not to a coding error in either of them.

The CMB lensing potential and the lensed spectra require the non-linear matter power spectrum, for which we use halofit \cite{Takahashi:2012em}. At parameter points where the interacting dark matter fraction is large and the dark sector decouples late, the linear matter power spectrum contains dark acoustic oscillations all the way up to the largest wavenumbers computed. At such points the extrapolation of the source functions beyond the largest computed wavenumber, which the non-linear module requires, can give a non-finite spectrum or an unphysical non-linear correction, and we have updated this extrapolation in \CLASSpp{} (Appendix~\ref{app:dao}). However, this does not make halofit (or HMcode) correct for spectra with dark acoustic oscillations. Neither of them has been calibrated with $N$-body simulations of such models, and ref.~\cite{Trendafilova:2025dce} shows that they cannot be relied upon there. We return to the size of the affected region of parameter space in Sec.~\ref{sec:results:params} and Appendix~\ref{app:dao}.

With the small-scale CMB data of ACT and SPT (Sec.~\ref{sec:numerics:data}) we use a later version of \CLASSpp{} with its precision settings for ACT DR6, except that the perturbations are solved up to $k = 10\,\mathrm{Mpc}^{-1}$ instead of $70\,\mathrm{Mpc}^{-1}$, which would make an evaluation prohibitively slow for this model (Appendix~\ref{app:dao}).

\subsection{Cosmological model: parameters and priors}
\label{subsec:param_and_prios}

The cosmological model consists of $\Lambda$CDM extended by two parameters for the self-interacting neutrino sector and four for DRMD, and the sampled cosmological parameters are
\begin{equation}
\begin{aligned}
    \theta = \{\,&\omega_\mathrm{b},\, \omega_\mathrm{cdm},\, H_0,\, \ln(10^{10}A_\mathrm{s}),\, n_\mathrm{s},\, \tau_\mathrm{reio},\, m_\mathrm{int},\,
                 \log_{10}(G_\mathrm{eff}\,\mathrm{MeV}^2),\\
                 &\Delta N_\mathrm{eff},\, f_\mathrm{idm},\, \log_{10} z_\mathrm{stop},\, \log_{10}(\mathcal{G}/\mathcal{H})_\mathrm{ini}\,\},
\end{aligned}
\end{equation}
twelve in total. The first six are the standard $\Lambda$CDM parameters, i.e.\ the physical densities of baryons and cold dark matter, the Hubble constant, the amplitude and spectral index of the primordial scalar power spectrum at $k = 0.05\,\mathrm{Mpc}^{-1}$, and the optical depth to reionisation. The neutrino sector consists of three degenerate self-interacting neutrinos which replace the three standard-model neutrinos, $m_\mathrm{int}$ is the sum of their masses (each has mass $m_\mathrm{int}/3$), and $G_\mathrm{eff}$ is the strength of the self-interaction. The neutrino temperature and a small massless relativistic component are set such that $N_\mathrm{eff} = 3.044$ in the absence of dark radiation. Finally, the DRMD sector is described by the dark radiation contribution $\Delta N_\mathrm{eff}$, the interacting fraction of the dark matter $f_\mathrm{idm}$, the redshift $z_\mathrm{stop}$ at which the Boltzmann suppression of the drag sets in, and the initial coupling $(\mathcal{G}/\mathcal{H})_\mathrm{ini}$.

We use uniform priors with the ranges listed in Table~\ref{tab:priors}. The ranges of $\omega_\mathrm{b}$, $H_0$ and $\ln(10^{10}A_\mathrm{s})$ extend to at least six standard deviations on either side of the posterior means. The ranges of the extended parameters follow the previous literature, except that $m_\mathrm{int}$, which is now the sum of the masses, lies in $[0.06, 0.3]$~eV, and that the self-interaction is limited to $\log_{10}(G_\mathrm{eff}\,\mathrm{MeV}^2) \leq -0.8$, beyond which the likelihood falls steeply. The lower limit of $m_\mathrm{int}$ is close to the smallest sum of the neutrino masses allowed by oscillation experiments, about 0.059~eV for normal ordering with a massless lightest neutrino \cite{Esteban:2024eli}. Furthermore, we require $f_\mathrm{idm} \leq 0.4$, which removes a narrow funnel towards $f_\mathrm{idm} = 1$ at vanishing $\Delta N_\mathrm{eff}$. Uniform priors on the logarithms of $G_\mathrm{eff}$, $z_\mathrm{stop}$ and $(\mathcal{G}/\mathcal{H})_\mathrm{ini}$ allow the sampler to scan their full physically interesting ranges. In addition we sample the nuisance parameters of the CMB likelihoods. For CamSpec these are three calibrations with Gaussian priors and six foreground parameters with uniform priors. With ACT and SPT (Sec.~\ref{sec:numerics:data}) the overall calibration of ACT is set equal to that of Planck, as in ref.~\cite{AtacamaCosmologyTelescope:2025blo}, and for SPT we use the priors of the collaboration's own combination with Planck and ACT \cite{SPT-3G:2025bzu}, among them a Gaussian prior on the temperature calibration. The internal priors of the SPT likelihood, among them one on $\tau_\mathrm{reio}$, are not used. In total we sample 21 parameters with Planck as the only CMB data and 24 with ACT and SPT, and with \HWCS{} also the absolute magnitude $M_B$ of the supernovae (Sec.~\ref{sec:numerics:data}).

We exclude the region $0 < \Delta N_\mathrm{eff} < 0.01$ from all posteriors, profile likelihoods and best fits. With so little dark radiation a large interacting fraction of the dark matter remains tightly coupled to it as a fluid with a small sound speed ($R \gg 1$, see Sec.~\ref{sec:theory:DRMD}), and this suppresses the matter power spectrum on small scales. The likelihood in this region therefore depends on the non-linear correction to such spectra, which is not calibrated for them and which we find to be numerically unreliable here (see Appendix~\ref{app:dao}). Note that $\Lambda$CDM and SI$\nu$, which have $\Delta N_\mathrm{eff} = 0$, are not affected. The emulators are trained over the whole prior, and the cut is applied to the posterior samples and to the searches for profile nodes and best fits. It removes 13\% of the \PDP{} posterior, 30\% of that of \HWC{} and less than $10^{-4}$ of the posteriors with SH0ES. In Sec.~\ref{sec:results:comparison} we compare the best fits inside this region with the ones we use.

\begin{table}[tb]
\centering
\begin{tabular}{lc}
\toprule
\textbf{Parameter} & \textbf{Prior range} \\
\midrule
$\omega_\mathrm{b}$ & $[0.0203,\ 0.0255]$ \\
$\omega_\mathrm{cdm}$ & $[0.079,\ 0.181]$ \\
$H_0\;[\mathrm{km\,s^{-1}\,Mpc^{-1}}]$ & $[57.8,\ 85]$ \\
$\ln(10^{10}A_\mathrm{s})$ & $[2.8,\ 3.28]$ \\
$n_\mathrm{s}$ & $[0.88,\ 1.06]$ \\
$\tau_\mathrm{reio}$ & $[0.01,\ 0.1]$ \\
$m_\mathrm{int}\;[\mathrm{eV}]$ & $[0.06,\ 0.3]$ \\
$\log_{10}(G_\mathrm{eff}\,\mathrm{MeV}^2)$ & $[-5.5,\ -0.8]$ \\
$\Delta N_\mathrm{eff}$ & $[0,\ 2]$ \\
$f_\mathrm{idm}$ & $[0,\ 0.4]$ \\
$\log_{10}z_\mathrm{stop}$ & $[2,\ 5]$ \\
$\log_{10}(\mathcal{G}/\mathcal{H})_\mathrm{ini}$ & $[2,\ 14]$ \\
\bottomrule
\end{tabular}
\caption{Uniform priors on the cosmological parameters. The prior is also the domain in which the emulators are trained. The results exclude the region $0 < \Delta N_\mathrm{eff} < 0.01$ (see text).}
\label{tab:priors}
\end{table}

\subsection{Datasets}
\label{sec:numerics:data}

We use two combinations of data, each with and without the SH0ES calibration, and all likelihoods are implemented in \cobaya{} \cite{Torrado:2020dgo}. The first combination, Planck+DESI+Pantheon+ (\PDP{}), consists of the following likelihoods:
\begin{itemize}
    \item \textbf{Planck}: the Planck 2018 low-$\ell$ temperature (Commander) and $E$-mode polarisation (SimAll) likelihoods \cite{Planck:2019nip}, the high-$\ell$ TT, TE and EE CamSpec likelihood built on the Planck PR4 (NPIPE) maps \cite{Rosenberg:2022sdy}, and the Planck PR4 CMB lensing reconstruction \cite{Carron:2022eyg}.
    \item \textbf{DESI BAO}: the baryon acoustic oscillation measurements of the second data release (DR2) of the Dark Energy Spectroscopic Instrument, from all tracers \cite{DESI:2025zgx}.
    \item \textbf{Pantheon+}: the Pantheon+ compilation of type Ia supernovae \cite{Brout:2022vxf}.
\end{itemize}
In \PDPS{} we replace the Pantheon+ likelihood by its \textit{Pantheon+SH0ES} version, in which the Cepheid distances of SH0ES \cite{Riess:2021jrx} to the host galaxies of the calibrating supernovae fix the absolute magnitude of the supernovae, and thereby the Hubble constant. In both versions the absolute magnitude is marginalised analytically.

The ground-based measurements of the CMB temperature and polarisation anisotropies by the Atacama Cosmology Telescope (ACT) \cite{AtacamaCosmologyTelescope:2025blo,AtacamaCosmologyTelescope:2025nti} and the South Pole Telescope (SPT) \cite{SPT-3G:2025bzu} now complement Planck in the damping tail, and both the neutrino self-interaction \cite{Kreisch:2019yzn} and the dark radiation of DRMD change this part of the spectrum. This data has changed the constraints on several of the proposed solutions to the Hubble tension \cite{Schoneberg:2026eys,Schoneberg:2026vaf}, although in the case of early dark energy ACT DR6 is less restrictive than Planck \cite{Poulin:2025nfb}. In the $H_0$ World Cup it is mainly the ACT data which constrains the models with additional dark radiation, among them self-interacting dark radiation and DRMD \cite{Schoneberg:2026vaf}, and a running of the spectral index relaxes these constraints \cite{Garny:2026gcs}. Our second combination is the baseline data combination of the $H_0$ World Cup \cite{Schoneberg:2026eys,Schoneberg:2026vaf} (\HWC{}), which allows a direct comparison of the SI$\nu$+DRMD model with the models ranked there, each of which contains a single new mechanism:
\begin{itemize}
    \item \textbf{Planck}: the low-$\ell$ temperature likelihood (Commander) as before, and the low-$\ell$ $E$-mode polarisation likelihood built on the SRoll2 maps \cite{Delouis:2019bub,Pagano:2019tci} in place of SimAll. The high-$\ell$ CamSpec PR4 likelihood \cite{Efstathiou:2019mdh,Rosenberg:2022sdy} is restricted to $\ell < 1000$ in TT and $\ell < 600$ in TE and EE. These are the cuts with which ACT combines its DR6 data with Planck \cite{AtacamaCosmologyTelescope:2025blo}, i.e.\ the small scales are left to the ground-based data.
    \item \textbf{ACT DR6}: the foreground-marginalised likelihood of the CMB-only TT, TE and EE bandpowers, ACT-lite \cite{AtacamaCosmologyTelescope:2025blo}, at $600 \leq \ell \leq 6500$.
    \item \textbf{SPT-3G D1}: the TT, TE and EE likelihood of the 2019 and 2020 observations of the SPT-3G Main field \cite{SPT-3G:2025bzu}, in its foreground-marginalised (``lite'') form \cite{Balkenhol:2024obe}, as implemented in candl \cite{Balkenhol:2024sbv}.
    \item \textbf{CMB lensing}: the ACT DR6 lensing power spectrum combined with that of Planck PR4 \cite{ACT:2023dou,ACT:2023kun,Carron:2022eyg}, in place of Planck PR4 lensing alone, and the SPT-3G lensing power spectrum of the 2019 and 2020 polarisation data, reconstructed with the MUSE method \cite{SPT-3G:2024atg}. Of the latter we use only the lensing part and not the delensed $EE$ spectrum of the same likelihood.
    \item \textbf{DESI BAO}: DESI DR2, all tracers, as before \cite{DESI:2025zgx}.
    \item \textbf{Pantheon+}: the Pantheon+ supernovae \cite{Scolnic:2021amr,Brout:2022vxf} at $z > 0.01$, uncalibrated, with the absolute magnitude marginalised analytically.
\end{itemize}
The Planck, ACT and SPT likelihoods are treated as independent, as in the joint analyses of the ACT and SPT collaborations \cite{AtacamaCosmologyTelescope:2025blo,SPT-3G:2025bzu}. In \HWCS{} the SH0ES calibration is added in the same way as in the $H_0$ World Cup, i.e.\ as a Gaussian prior on the absolute magnitude of the supernovae, $M_B = -19.253 \pm 0.027$ \cite{Riess:2021jrx}, not on $H_0$ \cite{Camarena:2021jlr}. The absolute magnitude is then sampled and not marginalised over. This differs from the Pantheon+SH0ES likelihood of \PDPS{}, which uses the Cepheid distances of the calibrating host galaxies directly, and the results of \PDPS{} and \HWCS{} therefore differ in the treatment of SH0ES as well as in the CMB data. Our analysis also differs from that of the $H_0$ World Cup in the version of the CamSpec likelihood, which is 12.6 in \cobaya{} and 12.7 there, in the non-linear power spectrum, for which the $H_0$ World Cup uses HMcode 2020 (see Sec.~\ref{sec:results:comparison}), and in the codes, since the $H_0$ World Cup uses \class{} with \montepython{} \cite{Audren:2012wb,Brinckmann:2018cvx} and the OL\'E emulator \cite{Gunther:2025xrq}.

\subsection{Likelihood emulation with \client{}}
\label{sec:numerics:client}

A Monte Carlo analysis of this model using \CLASSpp{} directly is not feasible. A single evaluation of the likelihood typically takes about 9 core seconds with \PDP{} and 70 core seconds with \HWC{}, the posterior is bimodal in $G_\mathrm{eff}$ (Sec.~\ref{sec:theory}), and several directions in parameter space are flat or almost flat. We therefore use \client{} \cite{Janken:2025wlq}, which emulates the likelihood itself and not the observables, as e.g.\ \connect{}~\cite{Nygaard:2022wri} does. For each dataset the emulators are trained on about $3\times10^5$ evaluations of the true likelihood, and the details of the training process can be found in Appendix~\ref{app:training}. \client{} is publicly available at \url{https://github.com/AarhusCosmology/client-f}.

\subsection{Parameter inference and model comparison}
\label{sec:numerics:inference}

The posterior distributions are obtained from the combined emulator of each dataset, inside the region made of ellipsoids around its boost points (Appendix~\ref{app:training}), using the nested sampler BEST \cite{Nygaard:2026fgl,Nygaard:2026hbr} with 2048 live points. Since the weight of each mode of a multimodal posterior varies between runs of a nested sampler with this number of live points, we combine four runs with independent seeds and estimate standard errors from the scatter between them. With \HWC{} the part of the posterior in the excluded region varies between 4\% and 48\%, and we combine 12 runs. Sampling the ellipsoids with an ensemble sampler instead gives the same results within the uncertainties, and the chains are analysed using GetDist \cite{Lewis:2019xzd}. We define the strongly interacting neutrino mode as the part of the posterior above the minimum of the marginalised posterior of $\log_{10}(G_\mathrm{eff}\,\mathrm{MeV}^2)$ between the two modes.

The models nested within the SI$\nu$+DRMD model lie on boundaries of its prior: $\Lambda$CDM ($\Delta N_\mathrm{eff} = f_\mathrm{idm} = 0$ and $\log_{10}(G_\mathrm{eff}\,\mathrm{MeV}^2) = -5.5$), SI$\nu$ ($\Delta N_\mathrm{eff} = f_\mathrm{idm} = 0$) and DRMD ($\log_{10}(G_\mathrm{eff}\,\mathrm{MeV}^2) = -5.5$). The lower boundary of the prior on $G_\mathrm{eff}$ is a proxy for vanishing self-interaction, since the likelihood is flat in $\log_{10} G_\mathrm{eff}$ well above it, and $m_\mathrm{int}$ is free in all four models. The best fit of a model is the point with the highest posterior found by the maximisations of the emulators, along the profile likelihoods and, for the nested models and for the SI$\nu$+DRMD model with \HWC{}, by minimisations with the true likelihood alone (Appendix~\ref{app:bestfits}). All quoted values of $\chi^2$ are calculated with the true likelihood.

Profile likelihoods are a well known complement to the posterior in cosmology \cite{Planck:2013nga,Herold:2024enb}, in particular where prior volume effects are important, as for early dark energy \cite{Herold:2021ksg}. Public codes such as CAMEL \cite{Henrot-Versille:2016htt}, PROSPECT \cite{Holm:2023uwa}, Procoli \cite{Karwal:2024qpt} and pinc \cite{Herold:2024enb} compute them by calling a Boltzmann solver, and they have also been computed with \connect{}, an emulator of the observables \cite{Nygaard:2023cus}. We compute one-dimensional profile likelihoods of all twelve cosmological parameters, with 25 nodes spanning the $\pm3\sigma$ range of the posterior and for some parameters the whole prior. At each node the emulator is maximised over all other parameters, and the result is refined with the true likelihood. With \HWC{}, where the profiles pass between two minima (Sec.~\ref{sec:results:profiles}), they were computed with the true likelihood alone (Appendix~\ref{app:bestfits}). We show the profiles as $\Delta\chi^2$ relative to the best fit and do not convert them into confidence intervals. The usual thresholds in $\Delta\chi^2$ give the correct coverage only if the likelihood is Gaussian in some parametrisation of the parameter, and this is not guaranteed here, in particular near the boundaries of the prior and for the bimodal self-interaction.

We compare the models using the difference in the Akaike information criterion \cite{Akaike:1974vps},
\begin{equation}
\Delta {\rm AIC} = \chi^2_{\rm model} - \chi^2_{\Lambda{\rm CDM}}  + 2 (N_{\rm model} - N_{\Lambda{\rm CDM}}),
\end{equation}
where $N$ is the number of free cosmological parameters ($N=7$, 8, 11 and 12 for $\Lambda$CDM, SI$\nu$, DRMD and SI$\nu$+DRMD). With \client{} the Bayesian evidence is not expensive to calculate, since it follows from the same nested sampling runs as the posteriors. A run of BEST evaluates the emulator about $2.7\times10^{8}$ times in one to one and a half hours on 16 cores, and with the true likelihood this would take about 76 core-years for \PDP{}, or 15 core-years for a sampler which needs five times fewer evaluations. In ref.~\cite{Sorensen:2025ywu} the evidences of inflationary models were calculated with the observables emulated by \connect{}, but with the likelihood still evaluated at all points, and BOBE \cite{Cohen:2026iij}, an emulator designed for the evidence, needs roughly a thousand times fewer evaluations of the likelihood than nested sampling.

Nevertheless, we compare models using the AIC instead of evidence ratios, because for this model the evidence ratios are determined by the prior volume and not by the data. Below $\log_{10}(G_\mathrm{eff}\,\mathrm{MeV}^2) \approx -3$ the neutrinos no longer interact on observable scales, and the likelihood is flat. The Bayes factor of the self-interaction against no self-interaction (with DRMD free in both cases) is then $\ln B = -\ln(W p_0)$, where $W = 4.7$ is the width of the prior and $p_0 = 0.11$ the posterior density at its lower boundary, and this gives $\ln B = 0.69 \pm 0.10$ (\PDP{}) and $0.68 \pm 0.14$ (\PDPS{}). With the prior restricted to the strongly interacting mode the same posterior would give 1.39 and 0.71, and as the lower boundary is moved towards vanishing coupling $\ln B$ tends to zero. The DRMD parameters behave in the same way as $\Delta N_\mathrm{eff} \to 0$. The AIC, on the other hand, does not depend on the prior ranges as long as the best fit lies inside them.

Here $\chi^2$ includes the Gaussian priors on the calibration parameters of the CMB likelihoods (Sec.~\ref{subsec:param_and_prios}), and with \HWCS{} also the prior on $M_B$. All other priors are uniform, and the best fits are thus the maxima of the posterior. Since $\chi^2$ is always calculated with the true likelihood, the emulators affect it only through the location of the best fits, and a point away from the true minimum can only overestimate the minimum. The searches with the true likelihood lowered the best fits of the emulators by at most 1.0 in $\chi^2$ for the SI$\nu$+DRMD model and by at most 2.4 for the nested models, except for DRMD with \HWC{}, whose best fit could only be found with the true likelihood (Appendix~\ref{app:bestfits}). We therefore estimate that the differences between the models are known to about $\Delta\chi^2 = 1$. Finally, the likelihood computed with \CLASSpp{} and \cobaya{} is not a perfectly smooth function of the parameters. Along short lines in parameter space it scatters about a smooth polynomial by about 0.01 in $\ln\mathcal{L}$ at the best fits and by up to 0.1 at the posterior means. This is negligible for the comparison of models, but it limits the refinement of the profile likelihoods (Appendix~\ref{app:bestfits}) and contributes to the measured error of the emulators (Appendix~\ref{app:accuracy}).

\section{Results}
\label{sec:results}

\subsection{Parameter inference results}
\label{sec:results:params}

\begin{figure}[tb]
\centering
\includegraphics[width=\textwidth]{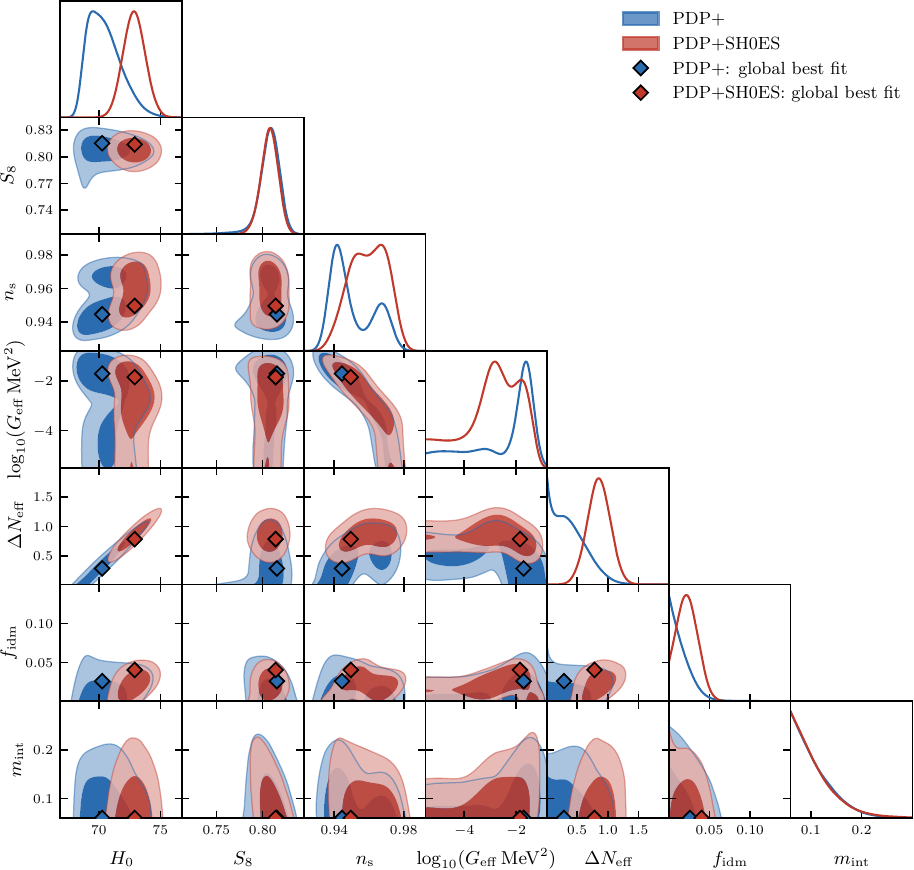}
\caption{Posterior distributions of the parameters that distinguish the model from $\Lambda$CDM, together with $H_0$, $S_8$ and $n_\mathrm{s}$, for the \PDP{} (blue) and \PDPS{} (red) datasets; 68\% and 95\% contours. Diamonds: the global best fits of the two datasets, the points with the highest posterior found (Sec.~\ref{sec:numerics:inference}). The axis of $f_\mathrm{idm}$ ends at 0.15, inside its prior (up to 0.4); beyond it lie 1.7\% of the \PDP{} posterior, a thin tail extending to the prior boundary, and $<0.1$\% of the \PDPS{} posterior.}
\label{fig:triangle_key}
\end{figure}

\begin{figure}[tb]
\centering
\includegraphics[width=\textwidth]{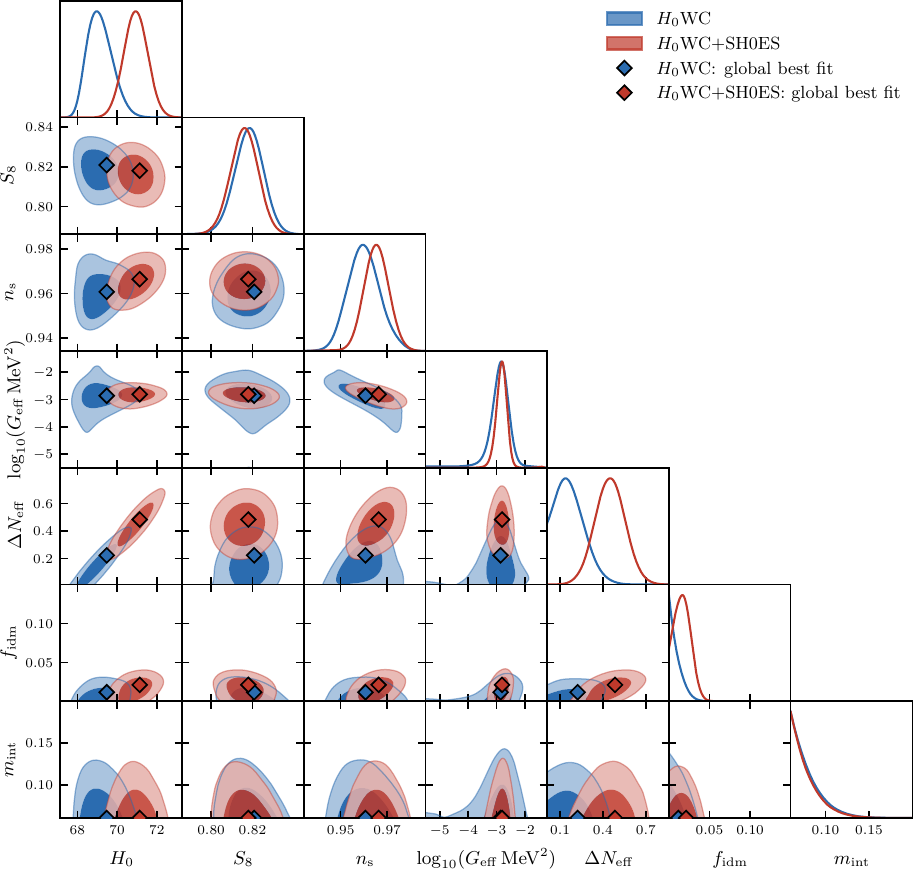}
\caption{As Fig.~\ref{fig:triangle_key}, for \HWC{} (blue) and \HWCS{} (red). Diamonds: the global best fits of the two datasets, the points with the highest posterior found. The axes of $f_\mathrm{idm}$ and $m_\mathrm{int}$ end at 0.15 and 0.2~eV, inside their priors; at most 0.1\% of either posterior lies beyond them.}
\label{fig:triangle_key_wc}
\end{figure}

The marginalised constraints on the cosmological parameters are listed in Table~\ref{tab:params} for the four datasets, and in Figs.~\ref{fig:triangle_key} and \ref{fig:triangle_key_wc} we show the posterior distributions of the parameters which distinguish the model from $\Lambda$CDM. The ground-based CMB data tightens most constraints considerably, and with \HWC{} the uncertainties of $\omega_\mathrm{b}$, $\omega_\mathrm{cdm}$ and $H_0$ are about half of those with \PDP{}.

\begin{table}[tb]
    \centering
    \renewcommand{\arraystretch}{1.1}
    \setlength{\tabcolsep}{3.5pt}
    \begin{tabular}{rlcccc}
    \cmidrule[\heavyrulewidth](lr){2-6}
    & & \multicolumn{2}{c}{\textbf{\PDP{}}} & \multicolumn{2}{c}{\textbf{\HWC{}}} \\
    \cmidrule(lr){3-4}\cmidrule(lr){5-6}
    & \textbf{Parameter} & without SH0ES & with SH0ES & without SH0ES & with SH0ES \\
    \cmidrule(lr){2-6}
    \multirow{6}{*}{\rotatebox[origin=c]{90}{$\Lambda$CDM}} & $10^2\omega_\mathrm{b}$ & $2.258^{+0.016}_{-0.021}$ & $2.281\pm 0.019$ & $2.250^{+0.012}_{-0.013}$ & $2.275\pm 0.013$ \\
     & $\omega_\mathrm{cdm}$ & $0.1269^{+0.0029}_{-0.0062}$ & $0.1355\pm 0.0039$ & $0.1214^{+0.0015}_{-0.0025}$ & $0.1267\pm 0.0024$ \\
     & $H_0\;[\mathrm{km\,s^{-1}\,Mpc^{-1}}]$ & $70.47^{+0.86}_{-1.7}$ & $72.87\pm 0.89$ & $69.15^{+0.52}_{-0.73}$ & $70.94\pm 0.60$ \\
     & $\ln(10^{10}A_\mathrm{s})$ & $2.998\pm 0.030$ & $3.012^{+0.031}_{-0.022}$ & $3.046\pm 0.016$ & $3.048\pm 0.014$ \\
     & $n_\mathrm{s}$ & $0.950^{+0.020}_{-0.015}$ & $0.960^{+0.012}_{-0.010}$ & $0.9601^{+0.0064}_{-0.0072}$ & $0.9654\pm 0.0055$ \\
     & $\tau_\mathrm{reio}$ & $0.0547^{+0.0067}_{-0.0077}$ & $0.0561^{+0.0069}_{-0.0078}$ & $0.0646^{+0.0058}_{-0.0069}$ & $0.0665^{+0.0059}_{-0.0068}$ \\
    \cmidrule(lr){2-6}
    \multirow{3}{*}{\rotatebox[origin=c]{90}{SI$\nu$}} & $m_\mathrm{int}\;[\mathrm{eV}]$ & $< 0.181$ & $< 0.189$ & $< 0.115$ & $< 0.113$ \\
     & $\log_{10}(G_\mathrm{eff}\,\mathrm{MeV}^2)$ & $-2.4^{+1.3}_{-1.1}$ & $< -1.43$ & $-2.91^{+0.36}_{-0.24}$ & $-2.82\pm 0.22$ \\
     & $P_\mathrm{strong}$ & $0.65 \pm 0.02$ & $0.30 \pm 0.02$ & $0.0094 \pm 0.0009$ & $0.0061 \pm 0.0006$ \\
    \cmidrule(lr){2-6}
    \multirow{4}{*}{\rotatebox[origin=c]{90}{DRMD}} & $\Delta N_\mathrm{eff}$ & $< 0.886$ & $0.85\pm 0.19$ & $< 0.352$ & $0.45\pm 0.10$ \\
     & $f_\mathrm{idm}$ & $< 0.0450$ & $< 0.0446$ & $< 0.0249$ & $< 0.0336$ \\
     & $\log_{10}z_\mathrm{stop}$ & $> 2.66$ & $> 3.20$ & $> 2.62$ & $> 3.37$ \\
     & $\log_{10}(\mathcal{G}/\mathcal{H})_\mathrm{ini}$ & unconstrained & unconstrained & unconstrained & unconstrained \\
    \cmidrule(lr){2-6}
    \multirow{5}{*}{\rotatebox[origin=c]{90}{Derived}} & $\sigma_8$ & $0.802^{+0.012}_{-0.0074}$ & $0.807^{+0.010}_{-0.0079}$ & $0.8144^{+0.0069}_{-0.0057}$ & $0.8180\pm 0.0063$ \\
     & $S_8$ & $0.807^{+0.012}_{-0.0081}$ & $0.8073^{+0.0097}_{-0.0086}$ & $0.8183\pm 0.0071$ & $0.8161\pm 0.0066$ \\
     & $\Omega_\mathrm{m}$ & $0.3033\pm 0.0042$ & $0.3003\pm 0.0041$ & $0.3029\pm 0.0037$ & $0.2986\pm 0.0036$ \\
     & $r_\mathrm{drag}\;[\mathrm{Mpc}]$ & $143.3^{+3.2}_{-1.6}$ & $139.0\pm 1.7$ & $145.8^{+1.3}_{-0.78}$ & $142.9\pm 1.1$ \\
     & $f_\mathrm{idr}$ & $0.048^{+0.021}_{-0.042}$ & $0.102^{+0.021}_{-0.019}$ & $0.0224^{+0.0090}_{-0.016}$ & $0.057\pm 0.013$ \\
    \cmidrule[\heavyrulewidth](lr){2-6}
    \end{tabular}
    \caption{Marginalised constraints on the parameters of the SI$\nu$+DRMD model for the four datasets: mean and 68\% interval, or the 95\% upper or lower limit where the posterior extends to the prior boundary. $P_\mathrm{strong}$ is the posterior probability of the strongly interacting neutrino mode, above the minimum of the posterior between the two modes (Sec.~\ref{sec:numerics:inference}), which lies at $\log_{10}(G_\mathrm{eff}\,\mathrm{MeV}^2) = -2.33$, $-2.19$, $-1.92$ and $-1.98$ for the four datasets. Its standard error is estimated from the scatter between independent runs of the nested sampler.}
    \label{tab:params}
    \renewcommand{\arraystretch}{1.0}
\end{table}

The most important difference between the two data combinations concerns the neutrino self-interaction. With Planck as the only CMB data the posterior of $\log_{10}(G_\mathrm{eff}\,\mathrm{MeV}^2)$ is bimodal, with a strongly interacting mode near $-1.6$ and a broad, moderately interacting mode at weaker coupling. The strongly interacting mode is correlated with a lower spectral index and a lower amplitude, and this degeneracy allows the mode in CMB data \cite{Kreisch:2019yzn,RoyChoudhury:2020dmd}. It contains $65 \pm 2$\% of the posterior for \PDP{} and $30 \pm 2$\% for \PDPS{}, and correspondingly the marginalised posterior of $n_\mathrm{s}$ is also bimodal. With the small-scale CMB data of ACT and SPT, on the other hand, the strongly interacting mode almost disappears. It contains only $0.9 \pm 0.1$\% of the posterior for \HWC{} and 0.6\% for \HWCS{}, and the posterior peaks at intermediate coupling, $\log_{10}(G_\mathrm{eff}\,\mathrm{MeV}^2) \approx -2.9$. Without SH0ES a tail towards weak interaction remains, whereas with SH0ES the posterior is a single narrow peak. This is in agreement with what was found using ACT DR6 \cite{AtacamaCosmologyTelescope:2025nti} and Planck, ACT and SPT combined \cite{Noriega:2026uhw}.

Without SH0ES the data does not require dark radiation, and we find $\Delta N_\mathrm{eff} < 0.886$ (\PDP{}) and $< 0.352$ (\HWC{}) at 95\%~CL. However, with SH0ES included dark radiation is detected. With \PDPS{} we find $\Delta N_\mathrm{eff} = 0.85\pm 0.19$ and $H_0 = 72.87\pm 0.89$~km\,s$^{-1}$\,Mpc$^{-1}$, in agreement with SH0ES, and with \HWCS{} $\Delta N_\mathrm{eff} = 0.45\pm 0.10$ and $H_0 = 70.94\pm 0.60$~km\,s$^{-1}$\,Mpc$^{-1}$. The damping tail measured by ACT thus limits the amount of extra radiation in the SI$\nu$+DRMD model, and thereby how much $H_0$ can be raised (Sec.~\ref{sec:results:profiles}). It should be noted that the two results also differ in the treatment of SH0ES itself (Sec.~\ref{sec:numerics:data}). In all four analyses the interacting fraction has only an upper limit of a few percent.

The diamonds in Figs.~\ref{fig:triangle_key} and \ref{fig:triangle_key_wc} mark the global best fit of each dataset, i.e.\ the point with the highest posterior found (Sec.~\ref{sec:numerics:inference}), which is also the reference point for the profile likelihoods. With \PDP{} and \PDPS{} both best fits lie in the strongly interacting mode, even though with SH0ES this mode contains only about a third of the posterior. The posterior mean, on the other hand, is not a good fit at all. At the mean of all 21 sampled parameters the true $\chi^2$ exceeds that of the best fit by $\Delta\chi^2 = 367$ for \PDP{} and $202$ for \PDPS{}, almost entirely in the high-$\ell$ CamSpec likelihood. The reason is that the mean combines parameter values from different parts of a multimodal and strongly non-Gaussian posterior, e.g.\ values of $n_\mathrm{s}$ and $A_\mathrm{s}$ averaged over the two modes. The marginalised constraints in Table~\ref{tab:params} therefore summarise the posterior distribution, but they do not describe a model which fits the data. For this purpose the best fits and the profile likelihoods are much better suited.

Above the cut at small $\Delta N_\mathrm{eff}$ (Sec.~\ref{subsec:param_and_prios}), the region of parameter space in which halofit is applied to spectra with dark acoustic oscillations (Sec.~\ref{sec:numerics:classpp}), roughly $f_\mathrm{idm} > 0.15$ with $z_\mathrm{stop} > 10^4$, contains 1.7\% of the \PDP{} posterior and very little of the other three. The uncalibrated non-linear modelling thus concerns only a small tail of the \PDP{} posterior (Appendix~\ref{app:dao}).

\subsection{Profile likelihoods}
\label{sec:results:profiles}

\begin{figure}[tb]
\centering
\includegraphics[width=\textwidth]{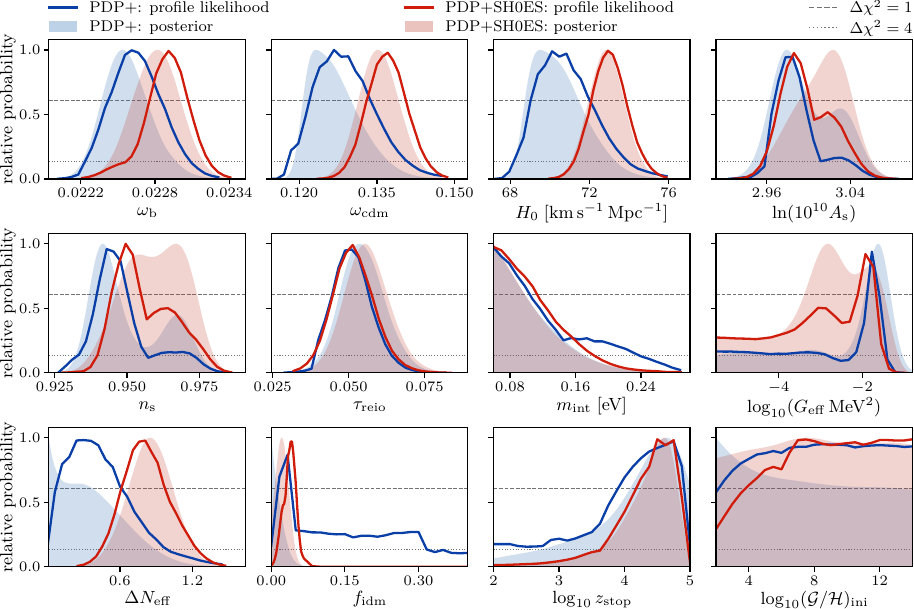}
\caption{Profile likelihoods (lines: the true likelihood computed with \CLASSpp{}, $e^{-\Delta\chi^2/2}$ relative to the best fit) and marginalised posteriors (shaded, normalised to a maximum of 1) of the twelve cosmological parameters, for \PDP{} (blue) and \PDPS{} (red). Where the profile and the posterior of a dataset differ, the posterior is affected by the volume of parameter space. The profile of $f_\mathrm{idm}$ for \PDPS{} continues at $\Delta\chi^2 \approx 20$ (relative probability $\sim 10^{-4}$) up to the prior boundary at 0.4. Dashed and dotted lines: $\Delta\chi^2 = 1$ and 4 ($e^{-1/2}$ and $e^{-2}$).}
\label{fig:profiles}
\end{figure}

\begin{figure}[tb]
\centering
\includegraphics[width=\textwidth]{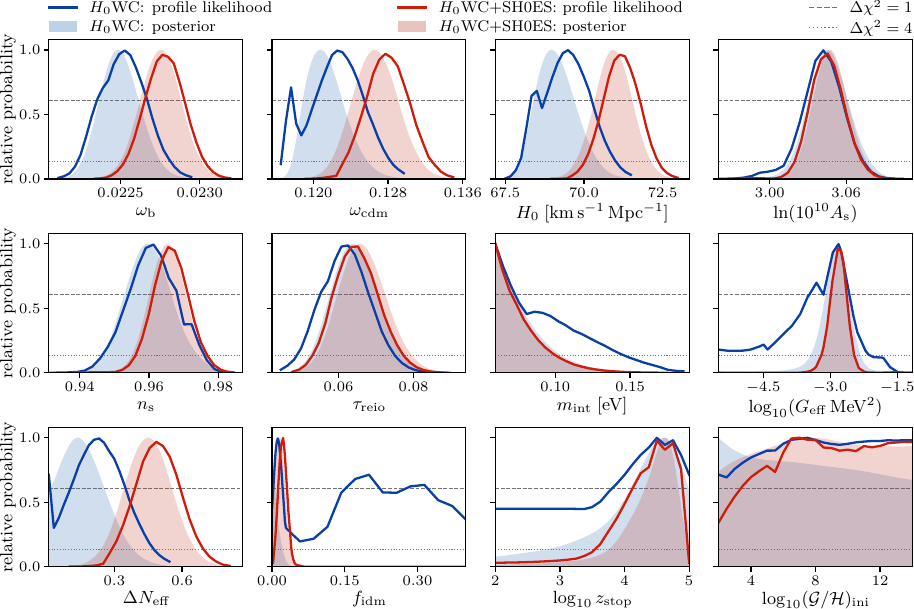}
\caption{As Fig.~\ref{fig:profiles}, for \HWC{} (blue) and \HWCS{} (red).}
\label{fig:profiles_wc}
\end{figure}

In Figs.~\ref{fig:profiles} and \ref{fig:profiles_wc} we compare the profile likelihoods with the marginalised posteriors of the same parameters. The differences between the two are mainly prior volume effects. The most important one is that the profiles prefer a non-zero interacting fraction, $f_\mathrm{idm} \approx 0.01$--0.04, whereas the marginalised posteriors give only upper limits. With SH0ES $f_\mathrm{idm} = 0$ is disfavoured by $\Delta\chi^2 = 9.6$ (\PDPS{}) and 6.9 (\HWCS{}), and without SH0ES by less. The reason is that the improvement requires $z_\mathrm{stop}$ in a narrow range, and it therefore occupies only a small part of the prior volume. Likewise, without SH0ES the \PDP{} profile prefers $\Delta N_\mathrm{eff} \approx 0.3$ and disfavours $\Delta N_\mathrm{eff} = 0$ by $\Delta\chi^2 = 3.6$, even though 16\% of the posterior lies at $\Delta N_\mathrm{eff} < 0.1$. In this region the three other DRMD parameters are unconstrained, and the plateau of the likelihood therefore has a very large volume.

The profiles of the self-interaction strength show the same change between the two data combinations as the posteriors. With \PDP{} and \PDPS{} the strongly interacting mode is the best fit, and the moderately interacting mode is worse by $\Delta\chi^2 = 3.6$ and $2.6$, respectively, although it retains a large part of the posterior. The contributions from the individual likelihoods (Appendix~\ref{app:breakdown}) show that the strongly interacting mode is driven by the high-$\ell$ Planck spectra, whereas the low-$\ell$ temperature spectrum prefers the moderately interacting mode. With CamSpec restricted to $\ell < 1000$ and the small scales measured by ACT and SPT this preference disappears. The best fit then lies at intermediate coupling, and in the strongly interacting mode the best fit of the SI$\nu$+DRMD model is worse by $\Delta\chi^2 = 4.3$ (\HWC{}) and $10.7$ (\HWCS{}). Without SH0ES the largest contributions to this difference are those of ACT-lite and the remaining CamSpec spectra, and with SH0ES that of SPT-lite (Table~\ref{tab:chi2_breakdown_wc}).

With SH0ES the contributions also show where the bounds on $H_0$ and $\Delta N_\mathrm{eff}$ come from. The supernovae calibrated by SH0ES set the lower bounds, and the upper bounds are set by CamSpec with \PDPS{} and by ACT-lite with \HWCS{}. SPT-lite constrains the dark radiation much less, and without SH0ES it even prefers slightly more of it. With the ground-based CMB data the SH0ES calibration and ACT-lite are thus in conflict, and at the best fit with SH0ES the $\chi^2$ of ACT-lite is higher than in $\Lambda$CDM (Table~\ref{tab:chi2_breakdown_wc}).

With \HWC{} the profiles pass between two minima. Besides the best fit there is a second minimum only 0.7 higher in $\chi^2$, at the boundary of the cut, $\Delta N_\mathrm{eff} = 0.01$, with a large interacting fraction, a weaker self-interaction and a smaller $H_0$ than the best fit, and with $z_\mathrm{stop}$ and $(\mathcal{G}/\mathcal{H})_\mathrm{ini}$ at the boundaries of their priors, i.e.\ in the region of dark acoustic oscillations (Appendix~\ref{app:dao}). This minimum lies below $\Delta\chi^2 = 1$ in the profiles of $\omega_\mathrm{cdm}$, $H_0$ and $\Delta N_\mathrm{eff}$ at their low ends, and in those of $f_\mathrm{idm}$ and $\log_{10}(G_\mathrm{eff}\,\mathrm{MeV}^2)$ (Fig.~\ref{fig:profiles_wc}). These parts of the profiles thus lie in a region where the non-linear modelling is not calibrated.

\subsection{Model comparison}
\label{sec:results:comparison}

The best-fit $\chi^2$ of the SI$\nu$+DRMD model, of DRMD, of SI$\nu$ and of $\Lambda$CDM are listed in Table~\ref{tab:model_comparison} for the four datasets, together with the differences in the AIC. Our most important result is that with the most aggressive data combination, \HWCS{}, the SI$\nu$+DRMD model fits the data much better than either SI$\nu$ or DRMD on its own. It lowers the $\chi^2$ of $\Lambda$CDM by 27.6 ($\Delta{\rm AIC} = -17.6$), whereas DRMD lowers it by 9.9 ($\Delta{\rm AIC} = -1.9$) and SI$\nu$ by 4.0 ($\Delta{\rm AIC} = -2.0$). Its $\chi^2$ is thus 17.7 lower than for DRMD, at the cost of one additional parameter ($\Delta{\rm AIC} = -15.7$), and its improvement over $\Lambda$CDM even exceeds the sum of the improvements of SI$\nu$ and of DRMD, 13.8. Most of the improvement over $\Lambda$CDM is due to the calibration itself, and the $\chi^2$ of the prior on $M_B$ drops from 31.4 in $\Lambda$CDM to 7.2 in the SI$\nu$+DRMD model. This is partly offset by the ground-based CMB data, in particular ACT-lite (Table~\ref{tab:chi2_breakdown_wc}).

\begin{table}[tb]
\centering
\begin{tabular}{lcccccccc}
\toprule
 & \multicolumn{4}{c}{\textbf{\PDP{}}} & \multicolumn{4}{c}{\textbf{\HWC{}}} \\
\cmidrule(lr){2-5}\cmidrule(lr){6-9}
 & \multicolumn{2}{c}{without SH0ES} & \multicolumn{2}{c}{with SH0ES} & \multicolumn{2}{c}{without SH0ES} & \multicolumn{2}{c}{with SH0ES} \\
\cmidrule(lr){2-3}\cmidrule(lr){4-5}\cmidrule(lr){6-7}\cmidrule(lr){8-9}
\textbf{Model} & $\Delta\chi^2$ & $\Delta$AIC & $\Delta\chi^2$ & $\Delta$AIC & $\Delta\chi^2$ & $\Delta$AIC & $\Delta\chi^2$ & $\Delta$AIC \\
\midrule
SI$\nu$+DRMD & $-9.13$ & $+0.87$ & $-37.56$ & $-27.56$ & $-5.50$ & $+4.50$ & $-27.59$ & $-17.59$ \\
DRMD & $-5.54$ & $+2.46$ & $-34.96$ & $-26.96$ & $-2.07$ & $+5.93$ & $-9.86$ & $-1.86$ \\
SI$\nu$ & $-5.49$ & $-3.49$ & $-12.99$ & $-10.99$ & $-2.39$ & $-0.39$ & $-3.96$ & $-1.96$ \\
\midrule
$\Lambda$CDM & \multicolumn{2}{c}{12393.20} & \multicolumn{2}{c}{12473.67} & \multicolumn{2}{c}{5765.85} & \multicolumn{2}{c}{5800.01} \\
\bottomrule
\end{tabular}
\caption{Best-fit $\chi^2$ of the four models relative to $\Lambda$CDM, $\Delta\chi^2$, and the difference in the Akaike information criterion, $\Delta$AIC, for the four datasets. The row of $\Lambda$CDM gives its $\chi^2$ itself. All values are calculated with the true likelihood at the best fits described in Sec.~\ref{sec:numerics:inference} and Appendix~\ref{app:bestfits}. The $\chi^2$ includes the Gaussian priors on the calibration parameters (Sec.~\ref{sec:numerics:inference}), and with \HWCS{} that of the prior on $M_B$. The numbers of free cosmological parameters are 12, 11, 8 and 7 for SI$\nu$+DRMD, DRMD, SI$\nu$ and $\Lambda$CDM.}
\label{tab:model_comparison}
\end{table}

The reason is clear from the best fits. With the self-interaction at intermediate coupling, the best fit of the SI$\nu$+DRMD model has more dark radiation and a larger Hubble constant than that of DRMD ($\Delta N_\mathrm{eff} = 0.48$ and $H_0 = 71.1$~km\,s$^{-1}$\,Mpc$^{-1}$, compared with $\Delta N_\mathrm{eff} = 0.31$ and $H_0 = 70.3$~km\,s$^{-1}$\,Mpc$^{-1}$), but it does not need the larger spectral index of DRMD. At this best fit the effect of the self-interaction on the TT, TE and EE spectra is largely reproduced, with the other parameters fixed, by a larger $n_\mathrm{s}$ and a positive running of the spectral index, $\alpha_\mathrm{s} \approx 0.02$, which relates our result to that of ref.~\cite{Garny:2026gcs}. The two differ mainly in the lensing potential, which is a few percent lower with the running. Most of the improvement over DRMD is due to the SH0ES calibration and CamSpec, with smaller contributions from the low-$\ell$ polarisation, the SPT lensing and ACT-lite, but the fit to the low-$\ell$ temperature spectrum is slightly worse (Table~\ref{tab:chi2_breakdown_wc}). This improvement is much larger than the uncertainty of the best fits, which is about $\Delta\chi^2 = 1$ (Sec.~\ref{sec:numerics:inference}). In particular, minimisations of DRMD with the true likelihood alone, started from four different points, converge to the same minimum within 0.09 in $\chi^2$ (Appendix~\ref{app:bestfits}).

In the other three cases the improvement over the individual models is smaller, and once the penalty for the additional parameters is taken into account the AIC does not significantly prefer the combination. With \PDPS{} the dark sector also gives a dramatically better fit than $\Lambda$CDM, but DRMD on its own accounts for most of it, with $\Delta\chi^2 = -35.0$ ($\Delta{\rm AIC} = -27.0$), almost entirely through the supernova likelihood. Adding the self-interacting neutrinos lowers $\chi^2$ by only 2.6 further, to $\Delta{\rm AIC} = -27.6$. Including the SH0ES calibration raises the best-fit $\chi^2$ of all models, since it includes the $\chi^2$ of the calibration data itself, but for DRMD and the SI$\nu$+DRMD model the increase is roughly 30 smaller than for $\Lambda$CDM. Within these models the tension between the SH0ES calibration and \PDP{} is thus much smaller than in $\Lambda$CDM. Without SH0ES the SI$\nu$+DRMD model lowers $\chi^2$ by 9.1 with \PDP{} and 5.5 with \HWC{}, but the AIC prefers it in neither case ($\Delta{\rm AIC} = +0.9$ and $+4.5$), and the same holds for DRMD.

\begin{figure}[tb]
\centering
\includegraphics[width=\textwidth]{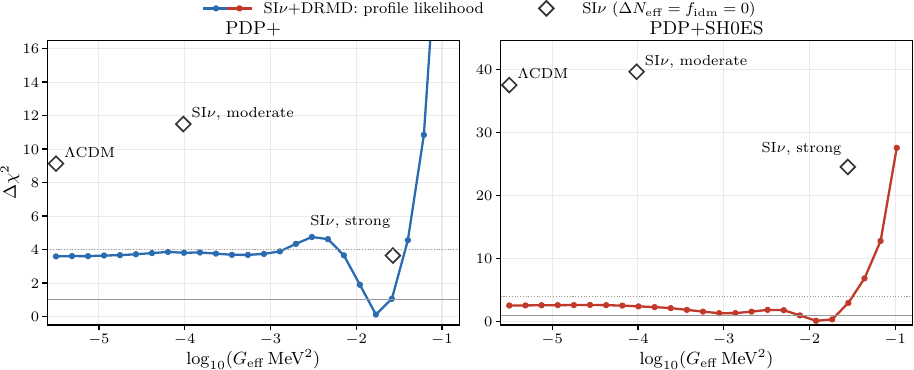}
\caption{The profile likelihood of the self-interaction strength in SI$\nu$+DRMD (the true $\chi^2$, refined, relative to the best fit of SI$\nu$+DRMD), and the true $\chi^2$ at the best fits on the boundary $\Delta N_\mathrm{eff} = f_\mathrm{idm} = 0$ (Sec.~\ref{sec:numerics:inference}): $\Lambda$CDM at $\log_{10}(G_\mathrm{eff}\,\mathrm{MeV}^2) = -5.5$, SI$\nu$ at $-4$ and its best fit in the strongly interacting mode. Horizontal lines: $\Delta\chi^2 = 1$ and 4.}
\label{fig:profile_logGeff}
\end{figure}

\begin{figure}[tb]
\centering
\includegraphics[width=\textwidth]{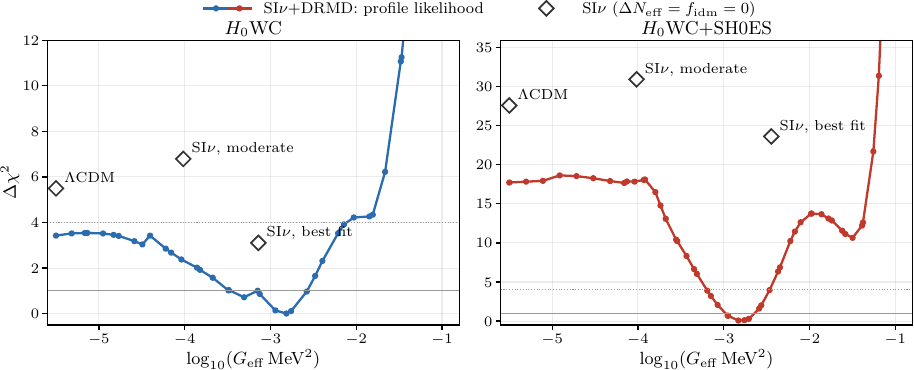}
\caption{As Fig.~\ref{fig:profile_logGeff}, for \HWC{} and \HWCS{}: the profile likelihood of the self-interaction strength over its whole prior, and the true $\chi^2$ at the best fits on the boundary $\Delta N_\mathrm{eff} = f_\mathrm{idm} = 0$ ($\Lambda$CDM, SI$\nu$ at $\log_{10}(G_\mathrm{eff}\,\mathrm{MeV}^2) = -4$, and the best fit of SI$\nu$; Sec.~\ref{sec:numerics:inference}).}
\label{fig:profile_logGeff_wc}
\end{figure}

With Planck as the only CMB data, self-interacting neutrinos on their own also fit the data clearly better than $\Lambda$CDM, with $\Delta\chi^2 = -5.5$ for \PDP{} ($\Delta{\rm AIC} = -3.5$) and $-13.0$ for \PDPS{} ($\Delta{\rm AIC} = -11.0$). The improvement comes entirely from the strongly interacting mode, and the moderately interacting mode does not give a lower $\chi^2$ than $\Lambda$CDM (Fig.~\ref{fig:profile_logGeff}). The strongly interacting mode is preferred by the high-$\ell$ CMB spectra and, without SH0ES, also by the DESI BAO, at the expense of the low-$\ell$ temperature spectrum. With SH0ES it also gives a better fit to the supernova data through a somewhat larger $H_0$. Once DRMD is included the two modes are much closer, since the extra radiation of DRMD has most of the effect of the strongly interacting mode, and this is also why the moderately interacting mode retains a large part of the posterior of the SI$\nu$+DRMD model. With ACT and SPT the best fit of SI$\nu$ lies outside the strongly interacting mode (Fig.~\ref{fig:profile_logGeff_wc}), and without SH0ES even the SI$\nu$+DRMD model lowers $\chi^2$ by only 1.2 in this mode (Table~\ref{tab:chi2_breakdown_wc}).

Our results can be compared directly with the ranking of the $H_0$ World Cup, which uses the same data, with the SH0ES calibration as a prior on $M_B$. In this ranking a model qualifies if $-\Delta{\rm AIC} > 10$ or $\ln{\rm BF} > 3$ relative to $\Lambda$CDM \cite{Schoneberg:2026eys}. Five models qualified, namely early dark energy, cold new early dark energy, Rock 'n' Roll, early modified gravity and a varying electron mass, the last two with $-\Delta{\rm AIC} = 19.6$ and 12.6. DRMD and self-interacting dark radiation were eliminated \cite{Schoneberg:2026eys}, and self-interacting neutrinos were not entered \cite{Schoneberg:2026vaf}. With $\Delta{\rm AIC} = -17.6$ the SI$\nu$+DRMD model would qualify by this criterion and would be ranked between early modified gravity and a varying electron mass. For DRMD we find $\Delta\chi^2 = -9.9$, compared with $-6.9$ in the $H_0$ World Cup. Most of this difference is due to the prior on the neutrino masses. Our prior, $\sum m_\nu \geq 0.06$~eV, is close to the lower bound from neutrino oscillations (Sec.~\ref{subsec:param_and_prios}), whereas that analysis uses $\sum m_\nu \geq 0$, and the best fits of both models lie at the lower boundary of our prior. If this boundary is moved to 0.003~eV, minimisations with the true likelihood give $\Delta\chi^2 = -7.9$ for DRMD, which differs from the result of the $H_0$ World Cup by about one. The difference from the $H_0$ World Cup is therefore a consequence of the preference of current data for a sum of the neutrino masses below the bound from oscillations \cite{Craig:2024tky,Green:2024xbb,DESI:2025zgx,Elbers:2025vlz}, and we keep the physical lower bound in our prior. The non-linear correction does not contribute to the difference, since using HMcode 2020, as that analysis does, instead of halofit changes the $\Delta\chi^2$ of DRMD by only 0.1. If DRMD is counted with three parameters, as in the $H_0$ World Cup where $(\mathcal{G}/\mathcal{H})_\mathrm{ini}$ is fixed \cite{Schoneberg:2026vaf}, it has at best $\Delta{\rm AIC} = -3.9$ and is still far from qualifying. The two analyses also differ in the version of CamSpec, the Boltzmann code and the methods used to find the best fits (Sec.~\ref{sec:numerics:data}). Finally, we have not computed the Bayes factors used in the ranking, for the reasons given in Sec.~\ref{sec:numerics:inference}.

Finally, the comparison hardly depends on the cut at small $\Delta N_\mathrm{eff}$ (Sec.~\ref{subsec:param_and_prios}), except for DRMD without SH0ES. In order to quantify the effect of the cut we have also minimised $\chi^2$ with the true likelihood inside the excluded region. With \HWC{} the best fit of the SI$\nu$+DRMD model there has a large interacting fraction of the dark matter which is tightly coupled to very little dark radiation, and a cosmology close to that of $\Lambda$CDM. Its $\chi^2$ is only 1.01 lower than at the best fit outside the region. For DRMD, on the other hand, the best fit with $\Delta N_\mathrm{eff} \geq 0.01$ lies at the boundary of the cut, in the region of dark acoustic oscillations, and inside the region DRMD lowers $\chi^2$ by 4.7 instead of 2.1. Without SH0ES the AIC does not significantly prefer any of the extensions, inside or outside the region. With \PDP{} the best fit of the SI$\nu$+DRMD model inside the region is worse than the one outside by $\Delta\chi^2 = 2.51$, and the best fit of DRMD lies outside the region. The cut thus does not affect our conclusions for \PDP{}. With SH0ES the region contains a negligible part of the posteriors. It should be stressed that these numbers depend on the uncalibrated non-linear modelling which led us to exclude the region, and they therefore indicate the potential effect of the cut, not a preference in the data (see also Appendix~\ref{app:dao}).

\FloatBarrier
\section{Discussion and conclusions}
\label{sec:conclusions}

Most of the extensions of $\Lambda$CDM which have been proposed to resolve the Hubble tension have been tested individually, and systematic comparisons rank the individual models as alternative solutions \cite{Schoneberg:2021qvd,Khalife:2023qbu,Schoneberg:2026eys,Schoneberg:2026vaf}. Combinations of such models have been studied much less, even though a single early-time mechanism may not be sufficient \cite{Vagnozzi:2023nrq}. In this paper we have studied one such combination as a worked example, namely self-interacting neutrinos together with the dark radiation matter decoupling of hot new early dark energy, two extensions which have each been proposed as a solution to the tension. We have analysed the model and the models nested within it using Planck, DESI and Pantheon+ (\PDP{}) and the more aggressive data combination of the $H_0$ World Cup (\HWC{}), in which the small-scale CMB spectra come from ACT and SPT, both with and without the SH0ES calibration.

Our most important result is that with the most aggressive data combination, \HWCS{}, the SI$\nu$+DRMD model fits the data much better than either SI$\nu$ or DRMD on its own. Its $\chi^2$ is 17.7 lower than for DRMD, at the cost of one additional parameter ($\Delta{\rm AIC} = -15.7$), and its improvement over $\Lambda$CDM exceeds the sum of the improvements of SI$\nu$ and DRMD on their own. These differences are much larger than the uncertainty of the best fits. The two extensions are thus complementary. At intermediate coupling the self-interaction allows more dark radiation and a larger Hubble constant than in DRMD, without requiring the larger spectral index which DRMD needs in order to compensate, and the posterior of the SI$\nu$+DRMD model gives $H_0 = 70.94\pm 0.60$~km\,s$^{-1}$\,Mpc$^{-1}$.

With the other three datasets this could not be established. In all four analyses the SI$\nu$+DRMD model reaches a lower best-fit $\chi^2$ than SI$\nu$ or DRMD, but with \PDPS{} DRMD on its own accounts for almost all of the improvement, and without SH0ES the AIC does not prefer the combination with either dataset. In summary, combined parameter spaces can contain better solutions than either individual model, but whether the improvement justifies the additional parameters depends on the data. Systematic comparisons of proposed solutions \cite{Schoneberg:2021qvd,Schoneberg:2026eys,Schoneberg:2026vaf} could very naturally be extended to include such combinations.

Such an analysis is not feasible when calling the Boltzmann code directly, since the SI$\nu$+DRMD model has twelve cosmological parameters, a bimodal posterior in the neutrino self-interaction and several flat directions in parameter space. With \client{} the emulators of a dataset are trained on about $3\times10^5$ evaluations of the true likelihood, and they reproduce the true log-likelihood at posterior samples with a root-mean-square error between 0.09 and 0.37. The posteriors and evidences were obtained from nested sampling of the emulators, and a single run would correspond to about 76 core-years of evaluations of the true likelihood. The profile likelihoods and the best fits were refined with a much smaller number of evaluations of the true likelihood, and all quoted $\chi^2$ values are thus those of the true likelihood. Models with many parameters, multimodal posteriors and flat directions are therefore within reach, and so are systematic studies of combined models.

In parameter spaces such as this one the marginalised posteriors are determined as much by the prior volume as by the likelihood. The profile likelihoods prefer a non-zero interacting fraction, whereas the posteriors give only upper limits, and the posterior mean is not a good fit at all. Similar prior volume effects have been found in profile likelihood studies of early dark energy \cite{Herold:2021ksg,Poulin:2025nfb}. For the same reason the evidence for the self-interaction depends on where the lower boundary of its logarithmic prior is placed, and we have compared models using their best fits and the AIC instead. The small-scale CMB data also changes the result for the self-interaction itself. The strongly interacting mode, which contains a large part of the posterior with Planck as the only CMB data, almost disappears with ACT and SPT, and an intermediate coupling is preferred, in line with refs.~\cite{AtacamaCosmologyTelescope:2025nti,Noriega:2026uhw}.

The main limitation of the analysis is the non-linear power spectrum of models with dark acoustic oscillations, for which neither halofit nor HMcode has been calibrated. We have excluded the region $0 < \Delta N_\mathrm{eff} < 0.01$, where the non-linear correction is also numerically unreliable, and above the cut the affected region contains only a small tail of the \PDP{} posterior. The best fits inside the excluded region indicate that the comparison of the SI$\nu$+DRMD model with $\Lambda$CDM hardly depends on the cut, whereas that of DRMD without SH0ES does. Whether the fits inside the region remain as good with a reliable non-linear modelling can only be settled by simulations of the non-linear evolution of such models. It would be very interesting to extend the comparison to other combinations of mechanisms, but this is beyond the scope of the present paper.

\section*{Reproducibility}

The analysis uses \client{} \cite{Janken:2025wlq}, the nested sampler BEST \cite{Nygaard:2026fgl,Nygaard:2026hbr}, \cobaya{} \cite{Torrado:2020dgo} and \CLASSpp{} with the SI$\nu$ and hot NEDE/DRMD implementations described in Sec.~\ref{sec:numerics:classpp}. The run settings, the \cobaya{} input files and the scripts which produce the tables and figures in this paper are publicly available, together with a description of how to reproduce the results, in the branch \texttt{papers/2026-sidr-drmd} of the \client{} repository: \url{https://github.com/AarhusCosmology/client-f/tree/papers/2026-sidr-drmd}.

\section*{Acknowledgements}
We thank Martin Sloth and Florian Niedermann for valuable comments on the manuscript. We acknowledge computing resources from the Centre for Scientific Computing Aarhus (CSCAA).

\appendix

\section{The \client{} emulators}
\label{app:client}

\subsection{Training}
\label{app:training}

In \client{} a neural network learns the log-posterior (i.e.\ the sum of the log-likelihoods and the log-prior) directly as a function of all sampled parameters, both cosmological and nuisance parameters (21 for \PDP{} and \PDPS{}, 24 for \HWC{} and 25 for \HWCS{}). The emulator is a fully connected network with five hidden layers of 512 neurons and the trainable activation function of ref.~\cite{SPECULATOR:2020}. It is trained on likelihood values computed with \CLASSpp{} and \cobaya{}, using a loss function which weights the training points by their proximity to the maximum of the likelihood.

The training points are acquired iteratively using active learning. The first 6000 points are drawn from a Latin hypercube in the prior volume. In the subsequent iterations the network is trained and then sampled with an affine invariant ensemble sampler \cite{Goodman:2010dyf,Foreman-Mackey:2012any} at a temperature $T = 7$ (i.e.\ the posterior raised to the power $1/T$), and 5000 new points are selected from the tempered chain. These points are chosen from a pool twenty times larger in order to fill regions where the training set is sparse, and a further 100 points are drawn around a quadratic fit to the best points found so far. Before a new chain is run the maxima of the network itself are checked against the true likelihood, so that the next training points are not drawn around maxima of the network which are not present in the true likelihood. The iterations continue until the scatter of the true minus the emulated log-likelihood over the newly acquired tempered points falls below $1.2\,T$ for two consecutive networks. For the two \PDP{} runs described below this happened after 9 and 7 iterations, and after 8 and 8 iterations for the two \PDPS{} runs.

The emulator is then refined with ten \emph{boosts}. Each boost draws 10\,000 points from the posterior of the current network at $T = 1$, divided equally between the posterior modes of the network (identified as groups of maxima between which a chain at $T=1$ does not mix), and thinned so that the training set follows the posterior density. After the first boost the network is trained on the deviation from a quadratic function fitted to the points near the maximum. Furthermore, it is used only inside a region made of ellipsoids around those boost points at which the true likelihood has been computed. Outside this region, where the network has not been trained on posterior points, the emulated likelihood is replaced by a large penalty.

The region of small $\Delta N_\mathrm{eff}$ is qualitatively different from the rest of parameter space. As $\Delta N_\mathrm{eff} \to 0$ the DRMD parameters $f_\mathrm{idm}$, $z_\mathrm{stop}$ and $(\mathcal{G}/\mathcal{H})_\mathrm{ini}$ become unconstrained and the likelihood develops a plateau, whereas at larger $\Delta N_\mathrm{eff}$ they are well constrained. We have therefore split the prior volume in $\Delta N_\mathrm{eff}$ and trained two emulators for each dataset, one on $0 \leq \Delta N_\mathrm{eff} \leq 0.13$ and one on $0.07 \leq \Delta N_\mathrm{eff} \leq 2$. After the boosts the two are combined into a single emulator in which each part is used only inside its own region, and in the overlap $0.08 < \Delta N_\mathrm{eff} < 0.12$ the two are blended with a weight whose first and second derivatives vanish at both ends of the band. Each of the four runs used a single node with 48 cores for 20--24 hours, and the two parts of an analysis were trained on roughly 299\,000 likelihood evaluations together.

The emulators of \HWC{} and \HWCS{} are trained with the same settings, including the split of the prior volume in $\Delta N_\mathrm{eff}$ and the blending of the two emulators of a dataset, and each of the four emulators (two datasets and two parts) is trained on 96 cores. The SPT lensing likelihood is undefined in a region of the prior volume with high $\omega_\mathrm{cdm}$ and $A_\mathrm{s}$. The reason is that it treats the lensing bandpowers as Gaussian in a transformed variable which exists only for bandpowers below roughly twice the fiducial ones. About 2\% of the points in the initial design fall in this region, far from the posterior, and they are excluded from the training set.

Derived parameters ($\sigma_8$, $S_8 = \sigma_8 (\Omega_\mathrm{m}/0.3)^{1/2}$, $\Omega_\mathrm{m}$, $r_\mathrm{drag}$ and the dark radiation fraction $f_\mathrm{idr}$) are emulated by a second network for each part, trained on the derived values computed at the same points. Its root-mean-square error at points not used in the training is at most 2.3\% of the spread of the parameter for $\sigma_8$ and $S_8$, and at most 0.7\% for $\Omega_\mathrm{m}$, $r_\mathrm{drag}$ and $f_\mathrm{idr}$. The decoupling redshift $z_\mathrm{dec}$ is not emulated reliably by the network, and we therefore do not report it.

\subsection{Accuracy}
\label{app:accuracy}

In contrast to an emulator of cosmological observables, the accuracy of a likelihood emulator can be measured directly in the region of high posterior probability. We draw 500 samples from the posterior of each combined emulator, compute the true log-likelihood at these points with \CLASSpp{} and \cobaya{}, and compare the two. Table~\ref{tab:accuracy} lists the root mean square and the largest absolute difference after one, five and ten boosts, for all four datasets. After ten boosts the root-mean-square difference is 0.14 (\PDP{}) and 0.09 (\PDPS{}), with mean differences of $+0.007$ and $+0.002$. This means that the emulator is accurate to well below one in $\ln\mathcal{L}$, well below the statistical width of the posterior. In fact part of this difference is not due to the emulator, but to the noise of the true likelihood itself (Sec.~\ref{sec:numerics:inference}). This noise is about 0.014 in $\ln\mathcal{L}$ at the best fits, and 0.10 (\PDP{}) and 0.06 (\PDPS{}) at the posterior means. If the noise is uncorrelated with the error of the emulators and no larger at the posterior samples than at the posterior means, subtracting the larger values in quadrature leaves 0.10 and 0.06 as a lower bound on the error of the emulators themselves.

In order to see why the error decreases with the number of boosts, we also compare the networks of the two parts with the true likelihood at the 10\,000 points of the following boost, i.e.\ at posterior samples on which they have not been trained. The first one or two boosts bring the largest improvement. The networks at the end of the iterative phase overestimate the likelihood at their own posterior samples, and for the part of \PDPS{} at larger $\Delta N_\mathrm{eff}$ the mean overestimate reaches 15 in $\ln\mathcal{L}$. The training on posterior samples corrects this. For the part at small $\Delta N_\mathrm{eff}$ the error converges to roughly 0.42 (\PDP{}) and 0.37 (\PDPS{}), and for the part at larger $\Delta N_\mathrm{eff}$ to roughly 0.11 for both datasets. The error of the combined emulator is mainly determined by whichever part covers the region containing most of the posterior, and with SH0ES the posterior lies entirely at $\Delta N_\mathrm{eff} > 0.1$.

\begin{table}[tb]
\centering
\caption{True minus emulated log-likelihood at 500 samples of the posterior of the combined emulator, after a given number of boosts: root mean square and largest absolute value.}
\label{tab:accuracy}
\begin{tabular}{lcccccc}
\toprule
 & \multicolumn{2}{c}{\textbf{after 1 boost}} & \multicolumn{2}{c}{\textbf{after 5 boosts}} & \multicolumn{2}{c}{\textbf{after 10 boosts}} \\
\cmidrule(lr){2-3}\cmidrule(lr){4-5}\cmidrule(lr){6-7}
\textbf{Data} & rms & largest & rms & largest & rms & largest \\
\midrule
\PDP{} & 0.74 & 4.95 & 0.26 & 1.68 & 0.14 & 1.53 \\
\PDPS{} & 0.34 & 1.77 & 0.15 & 0.85 & 0.09 & 0.64 \\
\midrule
\HWC{} & 1.34 & 14.58 & 0.98 & 9.07 & 0.37 & 2.74 \\
\HWCS{} & 0.31 & 1.61 & 0.16 & 1.14 & 0.11 & 0.71 \\
\bottomrule
\end{tabular}
\end{table}

For the emulators of \HWC{} the iterative phase ended after 7 and 7 iterations for the low and the high part, and for \HWCS{} after 6 and 7. For the parts at large $\Delta N_\mathrm{eff}$ the error converges in the same way as for \PDP{} and \PDPS{}, to 0.11 and 0.12 at the last boost. However, for the parts at small $\Delta N_\mathrm{eff}$ it converges more slowly, to 0.88 and 1.27. These errors are concentrated at $f_\mathrm{idm} > 0.1$, where $\Delta N_\mathrm{eff} \approx 0$ leaves the interacting fraction almost unconstrained. Along this flat direction the networks repeatedly produce new maxima at which they overestimate the likelihood, and these are corrected by the next boost. With SH0ES almost none of the posterior lies in this region.

\subsection{Dark acoustic oscillations and the non-linear power spectrum}
\label{app:dao}

Where a large fraction of the dark matter interacts with the dark radiation and decouples late, the logarithmic form $S(k) = S(k_\mathrm{max}) \ln(k/k_\ast)/\ln(k_\mathrm{max}/k_\ast)$ fitted to the last two points of the table, with which \CLASSpp{} extrapolates the source functions $S$ beyond $k_\mathrm{max} = 5\,\mathrm{Mpc}^{-1}$ (Sec.~\ref{sec:numerics:classpp}), becomes unstable. We found values of $\ln(k_\mathrm{max}/k_\ast)$ of e.g.\ $-1893$ and $+998$ at $z=0$, instead of a number of order unity. Without the regularisation the non-linear and lensed spectra were then returned as NaN at 10 of the points inside our prior at which we evaluated the likelihood, all of them at $\Delta N_\mathrm{eff} < 0.04$, $f_\mathrm{idm} > 0.2$ and $z_\mathrm{stop} > 10^{4.4}$, and raising $k_\mathrm{max}$ does not remove the problem. For \PDP{} and \PDPS{} we regularised the extrapolation at such points only, and we have verified that the spectra at random points of the prior volume and at the best fit are identical to those of the unmodified code. The later version of \CLASSpp{} used with \HWC{} and \HWCS{} (Sec.~\ref{sec:numerics:classpp}) instead continues the source functions with their asymptotic high-$k$ form, and it gives finite spectra at all the parameter points where the previous version failed. With the regularisation the spectra are finite, but the non-linear correction in this region is large. At eight points close to a good fit of the \PDP{} data halofit changes the lensing potential spectrum $C_L^{\phi\phi}$ by 5--13\% (3.3\% at that fit), which corresponds to $\Delta\ln\mathcal{L} = 1.2$--$6.6$ in the CamSpec and low-$\ell$ likelihoods alone. In other words, halofit makes points in this region fit the CMB better than linear theory does. At the points where the unregularised code failed the halofit $C_L^{\phi\phi}$ also changes by 0.8--6.8\% when $k_\mathrm{max}$ is raised from 5 to $30\,\mathrm{Mpc}^{-1}$. For the same reason we solve the perturbations with \HWC{} and \HWCS{} up to $k = 10\,\mathrm{Mpc}^{-1}$ and not up to the default $2\,\mathrm{Mpc}^{-1}$ of the later version, whose continuation assumes the high-$k$ form of cold dark matter. At points with dark acoustic oscillations $C_L^{\phi\phi}$ obtained with the default differs by up to 9\% at $L \leq 400$ from the one obtained with $10\,\mathrm{Mpc}^{-1}$, which in turn lies within 0.4\% of the result obtained with $30\,\mathrm{Mpc}^{-1}$.

Below $\Delta N_\mathrm{eff} \approx 0.01$ the non-linear correction is not only uncalibrated but also unstable. At a point with $\Delta N_\mathrm{eff} = 3\times10^{-4}$, $f_\mathrm{idm} = 0.15$ and decoupling at $z \approx 1100$, and relative to the same point with $f_\mathrm{idm} = 0$, the lensing potential spectrum at $L = 1000$ is 2\% lower in linear theory, unchanged with HMcode 2020 and 8\% higher with halofit. At $z = 0.5$ halofit raises the power at $k = 0.1\,\mathrm{Mpc}^{-1}$ by 19\% above the linear spectrum, compared with 1\% without the interacting dark matter. For this reason we exclude the region $0 < \Delta N_\mathrm{eff} < 0.01$ from the analysis (Sec.~\ref{subsec:param_and_prios}).

Above the cut the region $f_\mathrm{idm} > 0.15$, $z_\mathrm{stop} > 10^4$ contains 1.7\% of the \PDP{} posterior (Sec.~\ref{sec:results:params}), and in the \PDP{} profile of $f_\mathrm{idm}$ it is only $\Delta\chi^2 \approx 2.6$ worse than the best fit (Fig.~\ref{fig:profiles}). This is less than the uncertainty of the non-linear correction. Along the profiles the emulator overestimates $\ln\mathcal{L}$ by a few at isolated points of this region, which the refinement with the true likelihood corrects. However, at the 11 samples of the \PDP{} posterior in the region the true minus the emulated log-likelihood is only $+0.12$ on average (Appendix~\ref{app:accuracy}). With SH0ES the region is disfavoured by $\Delta\chi^2 \approx 20.1$. In summary, the small \PDP{} tail towards large $f_\mathrm{idm}$ at small $\Delta N_\mathrm{eff}$ depends on non-linear modelling which is not calibrated for this model, but the \PDPS{} results do not.

With \HWC{} the best fit of the SI$\nu$+DRMD model inside the excluded region (Sec.~\ref{sec:results:comparison}) is preferred by ACT-lite and, to a lesser extent, by the low-$\ell$ TT and EE spectra and DESI, whereas CamSpec, the lensing reconstructions of ACT and SPT, SPT-lite and the supernovae prefer the best fit outside the region. At the point inside the region $H_0$, $\omega_\mathrm{cdm}$ and $n_\mathrm{s}$ are all closer to their values in $\Lambda$CDM than to those at the best fit outside the region.

\section{Contributions of the individual likelihoods}
\label{app:supplementary}

\subsection{Profile likelihoods}
\label{app:breakdown}

\begin{figure}[tb]
\centering
\includegraphics[width=\textwidth]{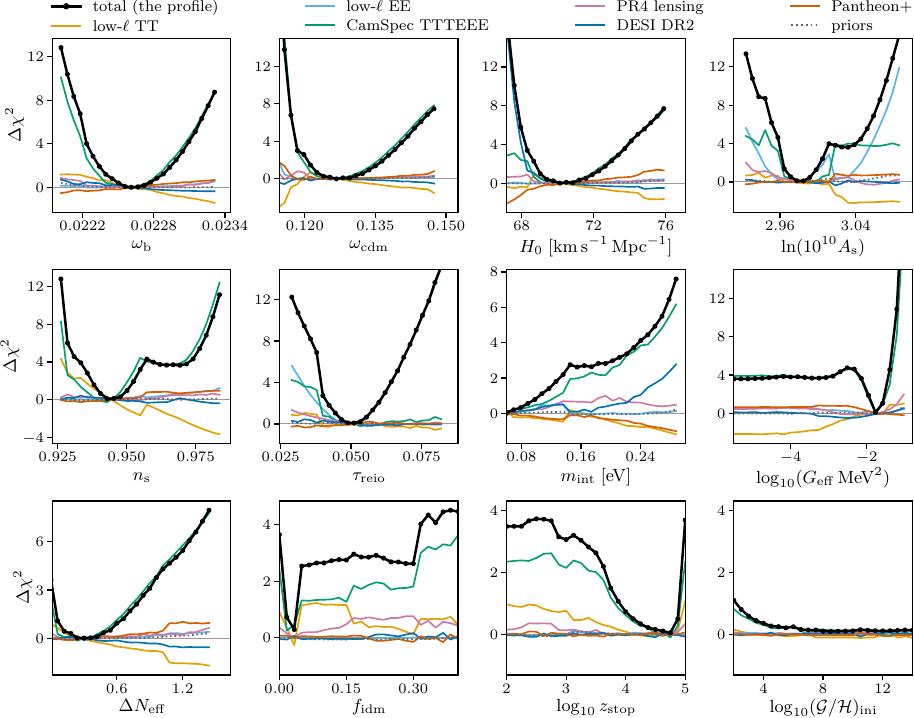}
\caption{The profile likelihoods of Fig.~\ref{fig:profiles} for \PDP{}, split into the contributions from the individual likelihoods: at the profile nodes, the $\chi^2$ of the individual likelihoods minus their values at the global best fit (coloured), and their sum (black), the profile's $\Delta\chi^2$. The priors are the Gaussian priors of the CamSpec calibrations. At $\Delta N_\mathrm{eff} = 0$ the values are those at the best fit of SI$\nu$, as in Fig.~\ref{fig:profiles}.}
\label{fig:profile_breakdown_pp}
\end{figure}

\begin{figure}[tb]
\centering
\includegraphics[width=\textwidth]{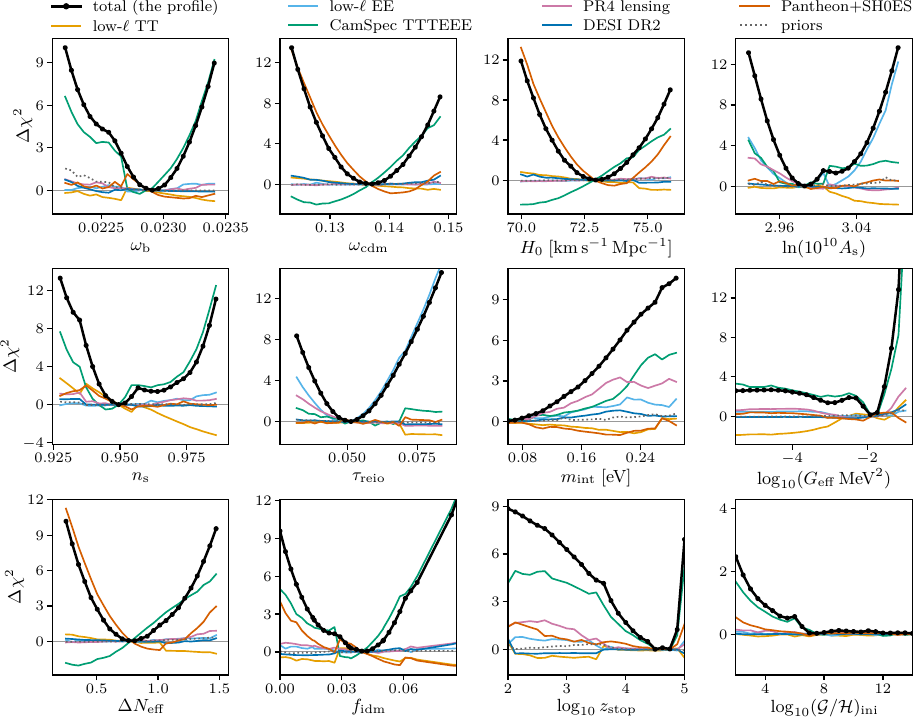}
\caption{As Fig.~\ref{fig:profile_breakdown_pp}, for \PDPS{}; the supernova likelihood includes the SH0ES calibration.}
\label{fig:profile_breakdown_sh}
\end{figure}

\begin{figure}[tb]
\centering
\includegraphics[width=\textwidth]{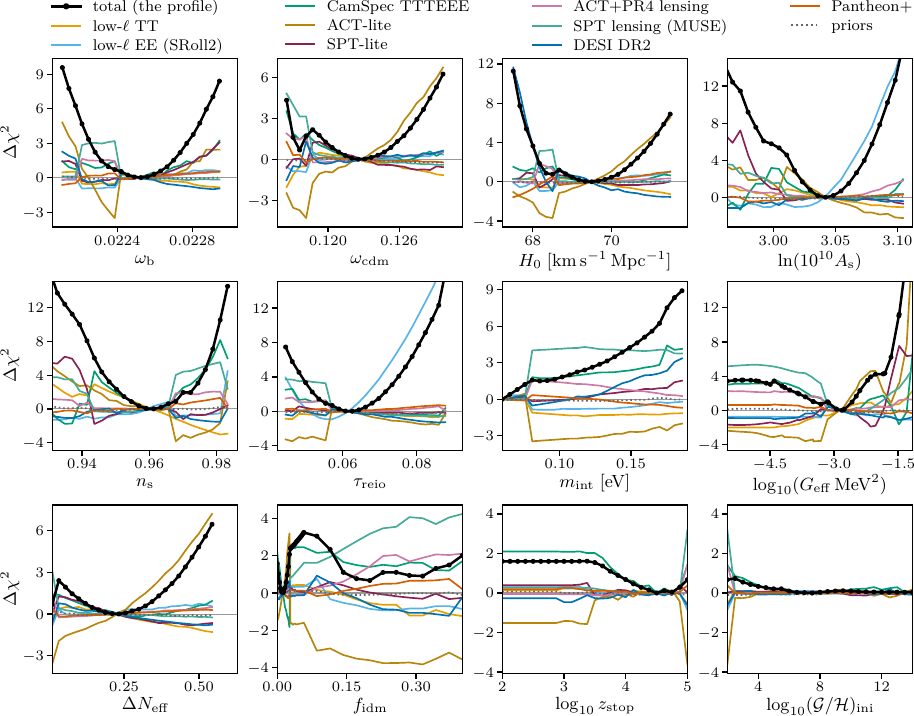}
\caption{As Fig.~\ref{fig:profile_breakdown_pp}, for \HWC{}; ``priors'' are those of the calibrations of CamSpec, ACT and SPT.}
\label{fig:profile_breakdown_wc}
\end{figure}

\begin{figure}[tb]
\centering
\includegraphics[width=\textwidth]{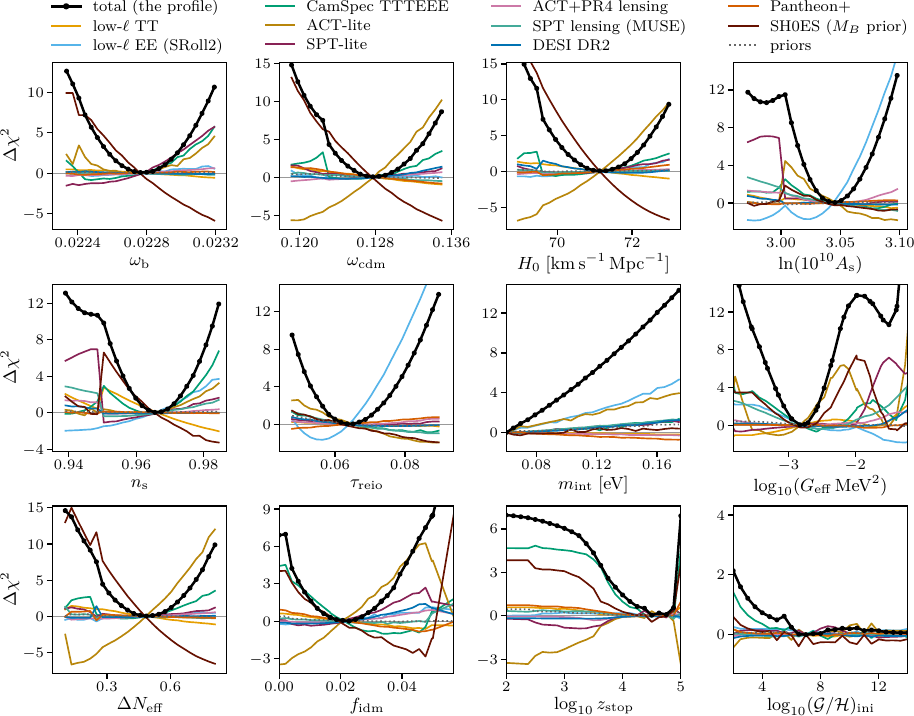}
\caption{As Fig.~\ref{fig:profile_breakdown_pp}, for \HWCS{}. The prior on $M_B$ is shown as the SH0ES calibration; ``priors'' are those of the calibrations of CamSpec, ACT and SPT.}
\label{fig:profile_breakdown_wcmb}
\end{figure}

All likelihoods have been computed at the nodes of the profile likelihoods. Figures~\ref{fig:profile_breakdown_pp} and \ref{fig:profile_breakdown_sh} split the profiles of Fig.~\ref{fig:profiles} into the contributions of the individual likelihoods, and Figs.~\ref{fig:profile_breakdown_wc} and \ref{fig:profile_breakdown_wcmb} do the same for those of Fig.~\ref{fig:profiles_wc}. The main features are discussed in Sec.~\ref{sec:results:profiles}. In addition, with \PDPS{} CamSpec prefers less dark radiation than the best fit has, and the increase in $\chi^2$ at $f_\mathrm{idm} = 0$ is due to the supernovae and CamSpec in roughly equal parts. Without SH0ES the upper bounds on $H_0$ and $\Delta N_\mathrm{eff}$ are again set by CamSpec, and the lower bound on $H_0$ by DESI. The optical depth and the amplitude are mainly determined by the low-$\ell$ polarisation together with CamSpec, and the neutrino mass by CamSpec with smaller contributions from the lensing and BAO likelihoods. With \HWCS{} the SH0ES calibration also sets the lower bound on $\omega_\mathrm{cdm}$, and ACT-lite the upper one. The increase in $\chi^2$ at $f_\mathrm{idm} = 0$ is due to CamSpec and SH0ES, and it is partly offset by ACT-lite. In the moderately interacting mode the largest contributions come from SH0ES and CamSpec. With \HWC{} the $\chi^2$ at $f_\mathrm{idm} = 0$ is higher by only 1.6, mainly because of CamSpec, and this is partly offset by ACT-lite.

\subsection{Best-fit $\chi^2$}
\label{app:chi2breakdown}

Table~\ref{tab:chi2_breakdown_wc} gives the $\chi^2$ differences to $\Lambda$CDM of the individual likelihoods at the best fits of the model comparison for \HWC{} and \HWCS{} (Sec.~\ref{sec:results:comparison} and Table~\ref{tab:model_comparison}).

\begin{table}[tb]
\centering
\begin{tabular}{rlrrrrr}
\cmidrule[\heavyrulewidth](lr){2-7}
& \textbf{Likelihood} & \textbf{SI$\nu$+DRMD} & \textbf{DRMD} & \textbf{SI$\nu$} & \textbf{SI$\nu$} & \textbf{SI$\nu$+DRMD} \\
& &  &  &  & (moderate) & (strong) \\
\cmidrule(lr){2-7}
\multirow{11}{*}{\rotatebox[origin=c]{90}{\HWC{}}} & low-$\ell$ TT & $+1.64$ & $-0.36$ & $+2.00$ & $+0.46$ & $+1.96$ \\
& low-$\ell$ EE & $-1.51$ & $-2.27$ & $-1.18$ & $+0.52$ & $-1.95$ \\
& CamSpec & $-3.81$ & $-0.82$ & $-2.04$ & $+0.39$ & $-2.06$ \\
& ACT & $+0.77$ & $-1.62$ & $-1.38$ & $-0.97$ & $+2.97$ \\
& SPT & $+1.21$ & $-0.46$ & $+2.62$ & $+0.34$ & $+1.19$ \\
& ACT+PR4 $\phi\phi$ & $-0.33$ & $+1.93$ & $-0.58$ & $-0.05$ & $+0.12$ \\
& SPT $\phi\phi$ & $-2.13$ & $+3.02$ & $-1.48$ & $-0.01$ & $-0.94$ \\
& DESI & $-1.17$ & $-2.07$ & $-0.13$ & $+0.59$ & $-2.01$ \\
& Pantheon+ & $+0.13$ & $+0.75$ & $-0.08$ & $-0.15$ & $-0.09$ \\
& Calibration priors & $-0.29$ & $-0.19$ & $-0.14$ & $+0.17$ & $-0.41$ \\
\cmidrule(lr){2-7}
& total & $-5.50$ & $-2.07$ & $-2.39$ & $+1.29$ & $-1.23$ \\
\cmidrule(lr){2-7}
\multirow{12}{*}{\rotatebox[origin=c]{90}{\HWCS{}}} & low-$\ell$ TT & $+0.63$ & $-0.84$ & $+3.58$ & $+0.53$ & $+1.09$ \\
& low-$\ell$ EE & $-2.45$ & $+0.26$ & $-3.51$ & $+0.27$ & $-4.03$ \\
& CamSpec & $-3.79$ & $+1.52$ & $+1.27$ & $+0.52$ & $-2.69$ \\
& ACT & $+6.38$ & $+8.22$ & $+1.02$ & $+0.09$ & $+6.30$ \\
& SPT & $-0.16$ & $-0.66$ & $+0.57$ & $-0.43$ & $+6.94$ \\
& ACT+PR4 $\phi\phi$ & $+0.07$ & $+0.48$ & $-0.33$ & $-0.02$ & $+1.23$ \\
& SPT $\phi\phi$ & $-2.75$ & $-0.52$ & $-2.79$ & $-0.12$ & $-0.65$ \\
& DESI & $-0.18$ & $-0.06$ & $-0.29$ & $+0.72$ & $+0.27$ \\
& Pantheon+ & $-0.57$ & $-0.29$ & $+0.35$ & $-0.35$ & $-0.64$ \\
& SH0ES & $-24.19$ & $-17.89$ & $-3.33$ & $+1.98$ & $-24.30$ \\
& Calibration priors & $-0.57$ & $-0.07$ & $-0.49$ & $+0.15$ & $-0.47$ \\
\cmidrule(lr){2-7}
& total & $-27.59$ & $-9.86$ & $-3.96$ & $+3.35$ & $-16.93$ \\
\cmidrule[\heavyrulewidth](lr){2-7}
\end{tabular}
\caption{The $\chi^2$ differences to $\Lambda$CDM of the individual likelihoods at the best fits of the models, for \HWC{} and \HWCS{}. ``SI$\nu$ (moderate)'' is the best fit of SI$\nu$ with $\log_{10}(G_\mathrm{eff}\,\mathrm{MeV}^2) = -4$, in the moderately interacting mode, and ``SI$\nu$+DRMD (strong)'' is the best fit of the SI$\nu$+DRMD model in the strongly interacting mode, above the minimum of the posterior between the modes (Sec.~\ref{sec:numerics:inference}). $\phi\phi$: the lensing likelihoods. SH0ES: the prior on $M_B$. Calibration priors: the Gaussian priors on the calibration parameters of the CMB likelihoods (Sec.~\ref{sec:numerics:inference}).}
\label{tab:chi2_breakdown_wc}
\end{table}

\FloatBarrier
\section{Best fits and profile likelihoods}
\label{app:bestfits}

The best fits of the nested models are first found by maximising the emulator of the SI$\nu$+DRMD model on their boundaries (Sec.~\ref{sec:numerics:inference}) with a quasi-Newton optimiser (L-BFGS-B), started from 60 points, i.e.\ the best training points and posterior samples placed on the boundary. Each of these best fits is then evaluated with \CLASSpp{} and \cobaya{}.

\subsection{\PDP{} and \PDPS{}}

The profile likelihoods are computed with 25 nodes per cosmological parameter spanning the $\pm3\sigma$ range of its posterior (Sec.~\ref{sec:numerics:inference}). At each node the emulator is first maximised over all other parameters, including the nuisance parameters, by basin hopping with a projected quasi-Newton ascent. The result is then refined using the true likelihood by a residual Newton method, in which the gradient of the difference between the true and the emulated log-likelihood is measured by finite differences, and steps are taken using the curvature of the emulator. The refinement required 27\,089 (\PDP{}) and 22\,984 (\PDPS{}) likelihood evaluations.

The refinement estimates gradients by finite differences, and because of the noise of the true likelihood (Sec.~\ref{sec:numerics:inference}) it often cannot establish that a node has converged. Only 149 of the 300 nodes for \PDPS{} and 88 for \PDP{} met the convergence criterion. The others ended when no step improved the likelihood further or when the maximum number of evaluations had been reached. For each node we use the best point found, and these points are consistent with smooth curves within the numerical noise, i.e.\ to a few tenths in $\Delta\chi^2$. The profiles span $\pm3\sigma$ of the posterior. For \PDPS{} this covers only small values of $f_\mathrm{idm}$, and its profile of $f_\mathrm{idm}$ was therefore extended over the whole prior in a separate computation. Wherever $\Delta\chi^2 < 4$ the profiles of the emulator itself agree with the true profiles to 0.05 (\PDPS{}) and 0.24 (\PDP{}) in $\Delta\chi^2$ (root mean square), in the latter case after a constant offset of 0.5.

For one node of the \PDP{} profile of $\omega_\mathrm{cdm}$ the maximum number of evaluations was reached while the value was still clearly off the smooth curve. This node was computed again with twice the limit, and we use the better of the two values. For \PDP{} 25 nodes had their optimum at $0 < \Delta N_\mathrm{eff} < 0.01$, and these were computed again with the searches restricted to $\Delta N_\mathrm{eff} \geq 0.01$. No node for \PDPS{} lay in this region. At the recomputed nodes we also took the best points of the neighbouring nodes and set the profiled parameter to the value of the node. Furthermore, we interpolated and extrapolated the optima of the neighbouring nodes along the profile to the node. The true likelihood was evaluated at all of these points, and at each recomputed node we use the best value found. Finally, at $\Delta N_\mathrm{eff} = 0$ for \PDP{} the refinement started from a maximum of the emulator which is not present in the true likelihood, and it reached only $\Delta\chi^2 = 9.5$. However, the best fit of SI$\nu$ (Sec.~\ref{sec:results:comparison}) lies on the same boundary with $\Delta\chi^2 = 3.6$, and we therefore use that value.

Where a point of the profiles lies on the boundary of a nested model and is better than the best fit of the emulator, we use it as the best fit of that model (Sec.~\ref{sec:numerics:inference}). This lowered the $\chi^2$ of the SI$\nu$+DRMD model by 0.6 (\PDP{}) and 0.8 (\PDPS{}), that of DRMD by 0.4 and 0.3, and that of SI$\nu$ by 0.6 with \PDP{}. As for \HWC{} and \HWCS{} (below), we have also minimised $\chi^2$ of $\Lambda$CDM, SI$\nu$ and DRMD with the true likelihood, started from the fits of the emulators after five and ten boosts and from the best points found by the other searches. For $\Lambda$CDM the three starts converge within 0.06 (\PDP{}) and 0.04 (\PDPS{}) in $\chi^2$, 0.3 and 0.5 below the best of the $\Lambda$CDM fits of the emulators. The minimum of SI$\nu$ is 0.5 and 0.7 below the best of its fits of the emulators, and that of DRMD lies within 0.01 of the point of the profiles on its boundary. These minima are the values used in Table~\ref{tab:model_comparison}.

\subsection{\HWC{} and \HWCS{}}

The best fit of each model is the point with the highest posterior (Sec.~\ref{sec:numerics:inference}) among the best fits of the emulators after one, five and ten boosts, each evaluated with \CLASSpp{}, the points of the profile likelihoods of the SI$\nu$+DRMD model and the results of minimisations with the true likelihood only. Such minimisations were made for $\Lambda$CDM, SI$\nu$ and DRMD with and without SH0ES, and for the SI$\nu$+DRMD model without SH0ES (see below). As in the rest of the paper, points at $0 < \Delta N_\mathrm{eff} < 0.01$ are excluded (Sec.~\ref{subsec:param_and_prios}). Since the models are nested, the best fit of a model is at least as good as that of any model it contains, and where none of its own fits is better we use that of the nested model. With these data the final emulators have maxima far from the posterior which are not present in the true likelihood (see the profile likelihoods below), and this was the case for the DRMD best fit of the final emulator without SH0ES. At the other best fits of the final emulators the true likelihood agrees with the emulators to within 0.9 (without SH0ES) and 0.6 (with SH0ES) in $\ln\mathcal{L}$.

With SH0ES the best fits found independently by the emulators after one, five and ten boosts agree to within 0.1 (SI$\nu$+DRMD) and 0.7 (DRMD) in $\chi^2$, and the true and emulated likelihoods agree at these points to within 0.06 in $\ln\mathcal{L}$. The best fit of the SI$\nu$+DRMD model in Table~\ref{tab:model_comparison} is a point of its profile likelihoods, and its $\chi^2$ is 1.0 lower than that of the best fit of the emulators.

With SH0ES we have in addition searched for the best fit of DRMD with the true likelihood instead of the emulator. For this we used a quasi-Newton search over the cosmological parameters with gradients from finite differences, and maximised over the nuisance parameters at all points. The search was started from four points, i.e.\ the best fits of the emulators, the best point found in the moderately interacting mode, and the best fit of the SI$\nu$+DRMD model with the self-interaction switched off. All four starts converge to the same minimum within 0.09 in $\chi^2$, 0.6 below the best of the DRMD fits of the emulators, and this is the value used in Table~\ref{tab:model_comparison}. The same minimisation for $\Lambda$CDM, started from the three $\Lambda$CDM fits of the emulators, gives a minimum 2.4 below the best of them, and this is the reference value for $\Delta\chi^2$ with SH0ES. The profile likelihood of $\log_{10}G_\mathrm{eff}$ in the moderately interacting mode, where the SI$\nu$+DRMD model reduces to DRMD, agrees with this minimum (Fig.~\ref{fig:profile_logGeff_wc}). We therefore estimate that the best fit of DRMD is known to well within $\Delta\chi^2 = 1$ and that of the SI$\nu$+DRMD model to about $\Delta\chi^2 = 1$, and the noise of the likelihood is only of order 0.01 (Sec.~\ref{sec:numerics:inference}). For SI$\nu$ the best fits of the three emulators differ by up to 2.5 in $\chi^2$, and by up to 7.5 without SH0ES. We have therefore also minimised $\chi^2$ of SI$\nu$ with the true likelihood, in the same way as for DRMD. The searches were started from these fits and from the strongly interacting mode, and without SH0ES also from the best point of the profile likelihoods on the boundary $\Delta N_\mathrm{eff} = f_\mathrm{idm} = 0$. The starts at intermediate coupling converge to the same minimum within 0.07 (with SH0ES) and 0.08 (without SH0ES) in $\chi^2$, 0.7 and 1.4 below the best of the SI$\nu$ fits of the emulators, and these are the values used in Table~\ref{tab:model_comparison}. The starts in the strongly interacting mode end 1.3 and 8.1 higher.

Without SH0ES the best fits of $\Lambda$CDM and DRMD were also found by minimisations with the true likelihood. For $\Lambda$CDM the minimisation was started from the three fits of the emulators, and it gives a minimum 0.5 below the best of them. For DRMD the fits of the emulators lie either inside the excluded region or at maxima which are not present in the true likelihood, and minimisations started near $\Lambda$CDM ended there. The best fit was instead found by a minimisation started from points of the profile likelihoods in the region of dark acoustic oscillations.

The profile likelihoods were computed in the same way as for \PDP{} and \PDPS{}, with 25 nodes per cosmological parameter spanning the $\pm3\sigma$ range of its posterior. In addition, the profiles of $f_\mathrm{idm}$ and $\log_{10}(G_\mathrm{eff}\,\mathrm{MeV}^2)$ cover the whole prior. For the latter the emulator was searched in its region (Appendix~\ref{app:training}) with all ellipsoids enlarged by a factor of two, since the boosts concentrate the training points on the dominant mode. Without SH0ES the searches of the emulator were restricted to $\Delta N_\mathrm{eff} \geq 0.01$ (see Sec.~\ref{subsec:param_and_prios}) by using a copy of the emulator run with its prior restricted to $\Delta N_\mathrm{eff} \geq 0.01$.

Along the profiles which cover the whole prior the final emulators have maxima far from the posterior which are not present in the true likelihood, in regions where they have not been trained by the boosts. At many nodes the $\chi^2$ of the emulator at its maximum is more than 100 lower than the $\chi^2$ calculated with the true likelihood at the same point. The refinement with the true likelihood then starts far from the profile, and at many of these nodes it reached the maximum number of evaluations. We have therefore taken the best points already computed, i.e.\ those of all other nodes and the best fits in Table~\ref{tab:model_comparison}, and set the profiled parameter to the value of a node. The true likelihood at such a point provides an upper bound on the profile at that node, and we use the lowest bound found. With SH0ES this changed 49 nodes.

Without SH0ES this was not sufficient, because the profiles pass between the two minima described in Sec.~\ref{sec:results:profiles}. For this dataset we have therefore computed the profiles with the true likelihood alone. At each node $\chi^2$ is minimised over the other parameters in the same way as for DRMD above, starting from the optimum of the previous node, and this is done along all profiles from both minima. For each node we use the best value found, and for most nodes this is lower than the value found by the refinement. This required about 116\,000 evaluations of the true likelihood, less than half of the 288\,000 used to train the emulators.

\FloatBarrier
\bibliographystyle{JHEP}
\bibliography{Bibliography/references}

\end{document}